# Perturbing Dynamics of Active Emulsions and Their Collectives

Muhammad Turab Ali Khan[1]*, Gaurav Gardi[1]*, Ren Hao Soon[1], Mingchao Zhang[1], Metin Sitti[1,2]†

[1] Physical Intelligence Department, Max Planck Institute for Intelligent Systems, 70569 Stuttgart, Germany.

[2] School of Medicine and College of Engineering, Koç University, 34450 Istanbul, Turkey.

* These authors contributed equally to this work.

† Corresponding author: sitti@is.mpg.de

**Abstract.**

Controlling fluidic flows in active droplets is crucial in developing intelligent models to understand and mimic single-celled microorganisms. Typically, these fluidic flows are affected by the interfacial dynamics of chemical agents. We found that these flows can be reconfigured by the mere presence of anisotropic solid boundary embedded within active droplets. Spontaneous fluidic flows dynamically orient an embedded magnetic cluster and the magnetic cluster, when realigned, causes these flows to reorient. Thus, providing an unprecedented control over the propulsion dynamics of chemotactic emulsions. When continuously perturbed, achiral emulsions exhibit emergent chiral motion with rotating fluidic flows. Such solid-fluid interactions removes barriers of specific emulsion chemistries and complements their inherent abilities thereby also enabling control over emergent collective behaviors of active droplets.

**Introduction:**

Single-celled microorganisms exhibit various intriguing behaviors, such as sensing, locomotion, and chemical communication. Active emulsions[1], such as oil droplets in water, are minimalistic model representations of such microorganisms[2]. Like single-celled microorganisms, these droplets exhibit chemotactic navigation[3,4] and propel via spontaneous symmetry breaking[5–7]. Chemotactic interactions among droplets lead to emergent collective behaviors[8–10], like their biological counterparts. However, key to understanding a droplet's behavior lies with controlling its spontaneous motion and being able to probe into inter-droplet chemotactic interactions without diminishing them, both of which precisely remain as a challenge[6]. These interactions and the motility are yielded by the fluidic flows, also known as Marangoni flows, within and outside the emulsions. The Marangoni flows are induced by spontaneous symmetry breaking of homogenous surfactant distribution across the liquid-liquid interface[2,11]. The spontaneous origin of these flows combined with the fluidic nature of emulsions makes it challenging to control their dynamics[6].

Here, we introduce a general strategy to mechanically perturb Marangoni flows within motile droplets independent of their chemical composition. We utilize a ferromagnetic nanoparticle cluster embedded within an oil droplet, as a responsive solid boundary to reconfigure the droplet's internal dynamics. Such reconfiguration of fluid flows allows for in-situ control of the droplet's chemotactic motion. While preserving the flows, our strategy can be applied in synergy with the interactions between droplets to control emergent collective behaviors of diverse emulsion systems. Overall, our results pave way for realizing controllable emulsion-based models for single-celled microorganisms and may help realize synthetic routes towards future autonomous and adaptive micromachines.

**Results:**

We devise oil droplets encapsulating small amount of ferromagnetic FePt nanoparticles in aqueous surfactant solutions (Fig. 1a). FePt is a strong ferromagnetic material (Supplementary Fig. 1) and it aggregates within oil droplets, forming a rod-like cluster of random morphology. To investigate the behavior of the magnetic cluster in the presence of Marangoni flows, we begin our discussion with 8CB (4-octyl-4′-cyanobiphenyl) droplets that demonstrate self-propulsion only above their smectic to nematic phase transition (at 33.5°C) and allow us to study cluster motion as the Marangoni flows establish (Supplementary Video 1). At low temperatures (below 33.5°C) or at low micellar concentrations (Supplementary Fig. 2), 8CB droplets and the cluster remain static with equilibrium polarized optical microscopy (POM) texture. Upon heating up, 8CB droplets dispersed in 10 wt% TTAB (Tetradecyltrimethylammonium bromide, 100 times critical micelle concentration (CMC)) begin to propel. Simultaneously, the magnetic cluster translates along the propulsion direction towards the frontal interfacial boundary while rotating about its longitudinal axis (Fig. 1b). Considering fluidic flows within spherical confinement, we expected the cluster to move freely, or even disperse, within the droplet under the influence of Marangoni flows[10,12]. However, we observe that magnetic cluster remains intact and maintains a specific orientation with respect to the fluidic flows and the angle between the cluster's longitudinal axis and the propulsion direction of the droplet remains bound and oscillates within ~60-90° (Fig. 1b).

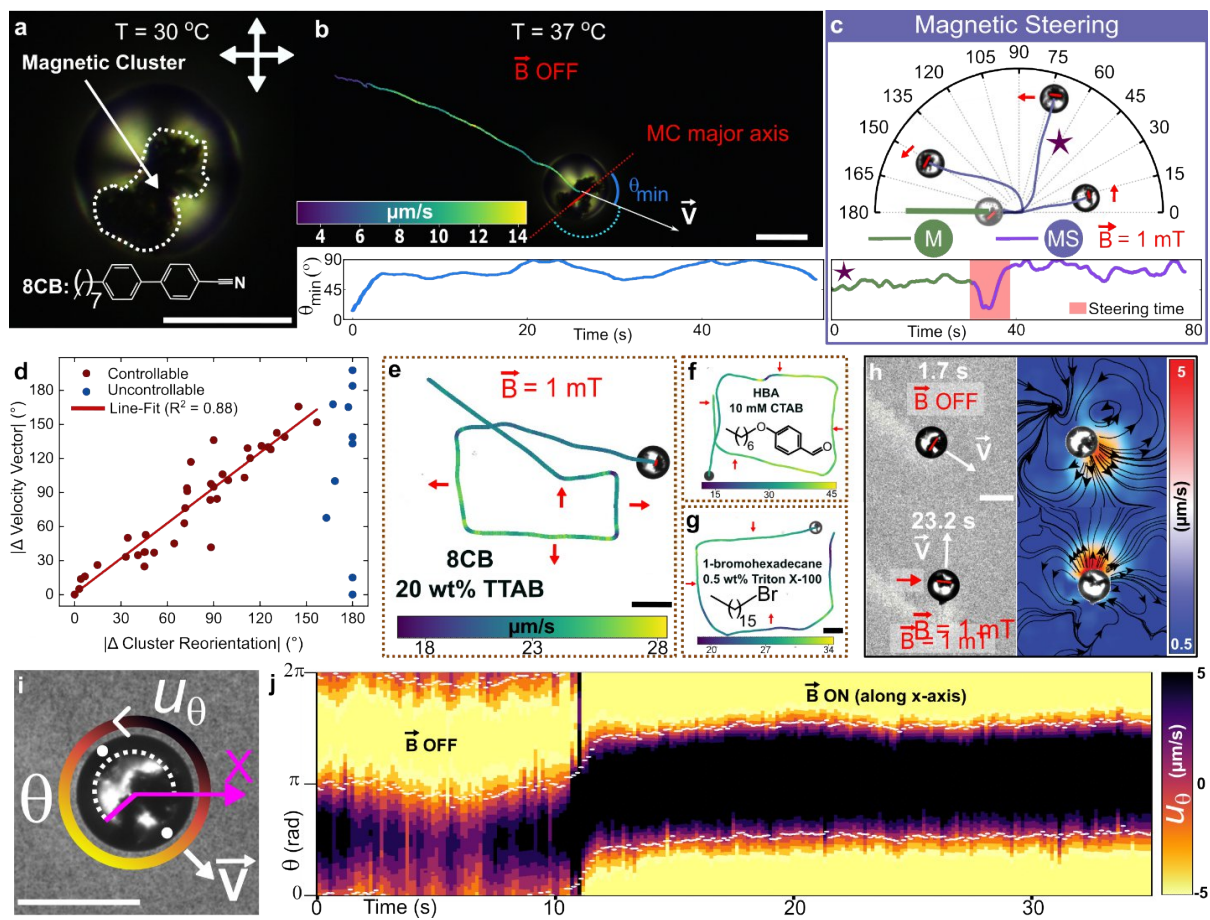

**Fig. 1. Steering self-propelling droplets encapsulating a cluster of ferromagnetic FePt nanoparticles.** (**a**) Cross-polarized image of a droplet in smectic state (T=30 °C) with a magnetic cluster at the center. (**b**) Representative trajectory of the droplet when heated to 37 °C (top). The angle between the magnetic cluster's longitudinal axis and the droplet's propulsion direction ($\theta_{min}$) remains bounded between ~60-90° (bottom). (**c**) Steering (MS, purple trajectory) self-propelling (M, green trajectory) droplets at 0°, 90° and 135° directions (top). $\theta_{min}$ temporarily reduces on application of magnetic field but quickly increases to ~90° (bottom). (**d**) Change in droplet's propulsion direction as a function of the extent of magnetic cluster rotation (47 data points), red (blue) dots correspond to cluster rotations below (beyond) 160°. (**e-g**) Patterning square trajectory using consecutive 90° cluster rotations under static fields (red arrow) with speed mapped over its trajectory for three different emulsion systems: (e) 8CB (4-octyl-4′-cyanobiphenyl), (f) HBA (4-Heptyloxybenzaldehyde) and (g) 1-bromohexadecane. (**h**) Bright-field images (left) and flow fields (right) in laboratory frame for a propelling droplet (velocity $\vec{V}$) before (top) and after (bottom) applying external magnetic field. Color bar represents the magnitude of flow velocity. (**i**) Schematic showing the tangential velocity ($u_\theta$) as a function of angle from x-axis ($\theta$) in the translating droplet frame. The color bar represents the value of $u_\theta$. (**j**) Kymograph showing the values of $u_\theta$ at different $\theta$ and time at a distance of 10 $\mu$m from droplet interface. The color bar represents the values of $u_\theta$; positive value means a counterclockwise flow and negative means clockwise flow. The white dots represent the two stagnation points ($\theta$ at which $u_\theta = 0$) at each time. Scale bars represent 100 $\mu$m.

Since the Marangoni flows dynamically orient the cluster perpendicular to the propulsion direction, we investigate whether fixing the orientation of the cluster could influence the Marangoni flows and the propulsion direction of the droplet. Thus, we fix the orientation of the cluster using an external magnetic field (B = 5mT) while the droplet is in smectic state (not propelling) and heat up until the droplets begin to propel. We notice that most of the droplets begin to propel randomly but gradually reorient to propel perpendicular to the aligned cluster (propulsion perpendicular to the applied field, Supplementary Fig. 3). Such adaption in the propulsion direction with respect to the cluster's orientation indicates that Marangoni flows can adapt and prefer to be oriented at a specific angle to the cluster's longitudinal axis. Moreover, the minimum angle distribution stays confined under magnetic field, further highlighting the orientational coupling between the cluster and the Marangoni flows (Supplementary Fig. 4).

Building on this finding, we wondered whether perturbing the orientation of the cluster while a droplet is propelling could reorient the Marangoni flows to influence the droplet's trajectory. We apply weak external static magnetic fields (B = 1 mT) to reorient the magnetic cluster in a propelling droplet. Figure 1c shows possible scenarios, where a self-propelling droplet, initially propelling in a straight-line (green), changes to a new trajectory (purple) under the influence of magnetic field at different angles (0º, 90º and 135º). The cluster quickly rotates (counterclockwise) inside the droplet to align along the magnetic field, while the droplet slowly rotates along the direction of cluster rotation and adapts to a new trajectory (purple) perpendicular to the magnetic cluster's orientation. The minimum angle evolves over time and captures the transition between self-propelling droplet and the magnetically steered droplet (Fig. 1c). Furthermore, the lag between the reorientation of magnetic cluster and the self-propelling droplet results in a finite curvature of the new trajectory (purple) and this lag depends on the extent of rotation of the magnetic cluster (Supplementary Fig. 5). The lag time also increases with decreasing cluster size (Supplementary Fig. 6). Additionally, the extent of change in the propulsion direction of self-propelling droplets linearly varies with the extent of change in the cluster's orientation (red dots in Fig. 1d).

Beyond 160º magnetic cluster rotation, a nonlinearity arises in the trend (blue dots in Fig. 1d and Supplementary Fig. 7) and steering for such large rotations becomes uncontrollable, as the direction of propulsion is not deterministic. We reason that for such large rotations, the self-propelling droplets interact with their own repulsive solubilized trails at the rear[13,14]. Therefore, to minimize the repulsive interactions with the filled micelles, the droplets propel in different directions. Although, our experimental design provides isotropic conditions (see Materials and Methods section 'Magnetic steering and curling'), the solubilized oil trail creates a local anisotropy, limiting the steering window of self-propelling droplets. Nevertheless, the self-propelling droplets can still be steered beyond this upper limit by using two or more consecutive magnetic cluster rotations, as demonstrated by the patterned square trajectory (Fig. 1e and Supplementary Video 1). Further, we probe cluster's orientation and feasibility of such magnetic steering in other Marangoni flow-driven droplet systems (isotropic oils: HBA (4-Heptyloxybenzaldehyde) and 1-bromohexadecane). Even changing the chemical structures of oils and the surfactants responsible for self-propulsion did not affect the cluster's dynamic assembly and different emulsion systems could be steered by changing the orientation of the cluster (Fig. 1f and g). Because the magnetic steering is not limited to a specific emulsion system, droplet size (Supplementary Figs. 8 and 9) and cluster size, we hypothesize that the magnetic cluster acts as a solid boundary and directly influences the Marangoni flows within the droplets.

To further probe into the physical effect of cluster's orientation on the Marangoni flow profiles, we first investigate internal flow structure of the droplet by taking advantage of POM texture of 8CB droplets. In a propelling droplet, point defect orients towards the propulsion direction (Supplementary Fig. 10) and is an indicator for the internal flow structure[6,15]. We notice that POM texture of the cluster-doped droplets varies from bare 8CB droplet (Supplementary Fig. 10b and c), likely due to the distortions in the nematic field lines caused by the presence of the cluster and other optical limitations[16,17]. Nevertheless, the propulsion direction and the internal flow structure changes with the cluster's orientation (Supplementary Fig. 10c). We quantify the fluid flows around the droplets because external and internal flow fields are coupled. Figure 1h shows the Marangoni flows around the droplet (in the laboratory frame of reference) and changing the orientation of the cluster leads to a reorientation of the Marangoni flow fields (Supplementary Video 2). To observe the evolution of the flow fields as we steer the droplet, we investigate the tangential flow velocity ($u_\theta$) around the droplet over time (Fig. 1i and j, in the droplet frame of reference). $u_\theta$ follows a sinusoidal-like profile with flow values zero at front and the rear of the droplet corresponding to the two stagnation points. The two bands in tangential flow velocity ($u_\theta$) correspond to a steady flow field with a dipolar symmetry[18]. As we perturb the cluster, a strong rotational flow arises visible as a monochromatic strip in kymograph (~ t = 11s), resulting in a sinusoidal-like profile with an offset (Supplementary Fig. 11). The cluster stops rotating once it aligns along the static fields, causing the rotational flows to vanish quickly. Concurrently, the Marangoni flows slowly adapt to the new orientation of the cluster (Supplementary Fig. 12), resulting in the finite steering time of the droplet's trajectory.

Static fields hold the cluster still while the Marangoni flows slowly adapt and slow, consecutive reorientations of the cluster can be used for stepwise steering of the droplet. Next, we study droplet's propulsion dynamics when the cluster is reoriented faster than the time required by the Marangoni flows to fully reorient. Therefore, we utilize rotating magnetic fields to continuously rotate the cluster in a self-propelling droplet. Unlike static fields, a rotating cluster perpetually perturbs the Marangoni flows resulting in a curling motion of the droplet (Fig. 2a). Concurrently, the dipolar symmetry axis of the tangential flow velocity starts to rotate across the droplet's interface, as indicated by a positive slope of the $u_\theta$ bands in the kymograph (Fig. 2b). At any given time, the tangential flows around the droplet are anisotropic at 0.1 Hz (Fig. 2c), indicating a co-existence of rotational and Marangoni flows as the droplet curls. While as the frequency of rotation increases, such anisotropy dies down (Figs. 2d-f), because the magnitude of the rotational flows scales with the applied frequency of the magnetic field (Fig. 2g) but the magnitude of Marangoni flows remains the same.

Consequently, we begin to notice observable differences in the droplet's trajectory as the frequency increases (Supplementary Fig. 13 and Supplementary Videos 3 and 4). We quantify these differences via mean-squared displacement (MSD) analysis (Fig. 2h)[15]. Curling motion is indicated by a dip in the plot of MSD vs time lag ($\tau$), as opposed to the monotonously increasing MSD for a straight-line trajectory (MSD $\propto \tau^2$). We also observe that the position of the dip changes with the frequency of the rotating magnetic field and it corresponds to the period of oscillations in the angular autocorrelation of curling droplet's velocity (Supplementary Fig. 14). At higher rotation frequency, the rotational dynamics become faster than the translation dynamics, thus decreasing the effective displacement of the droplet in one rotation. The average speed decreases with increasing the frequency of the applied magnetic field, and the droplet almost comes to a halt at 1 Hz (Fig. 2i). The amplitude of the error bars shows the magnitude of oscillations in the speed and the oscillation frequency increases with increasing the frequency of the magnetic cluster rotation (Supplementary Figs. 15 and 16). Such frequency-based response indicates that at lower frequencies, a rotating cluster can continuously but not synchronously reorient the Marangoni flows. While at higher frequencies, faster rotational flows dominate and overwhelm the Marangoni Flows. Typically, droplets undergo chiral motion due to inherent instability[15,19] or presence of chiral agents[20]. The chiral trajectory of our droplets is generated by perturbing the Marangoni flows, thus allowing for controllable switching between chiral and straight-line trajectories (Supplementary Fig. 13 and Supplementary Video 3).

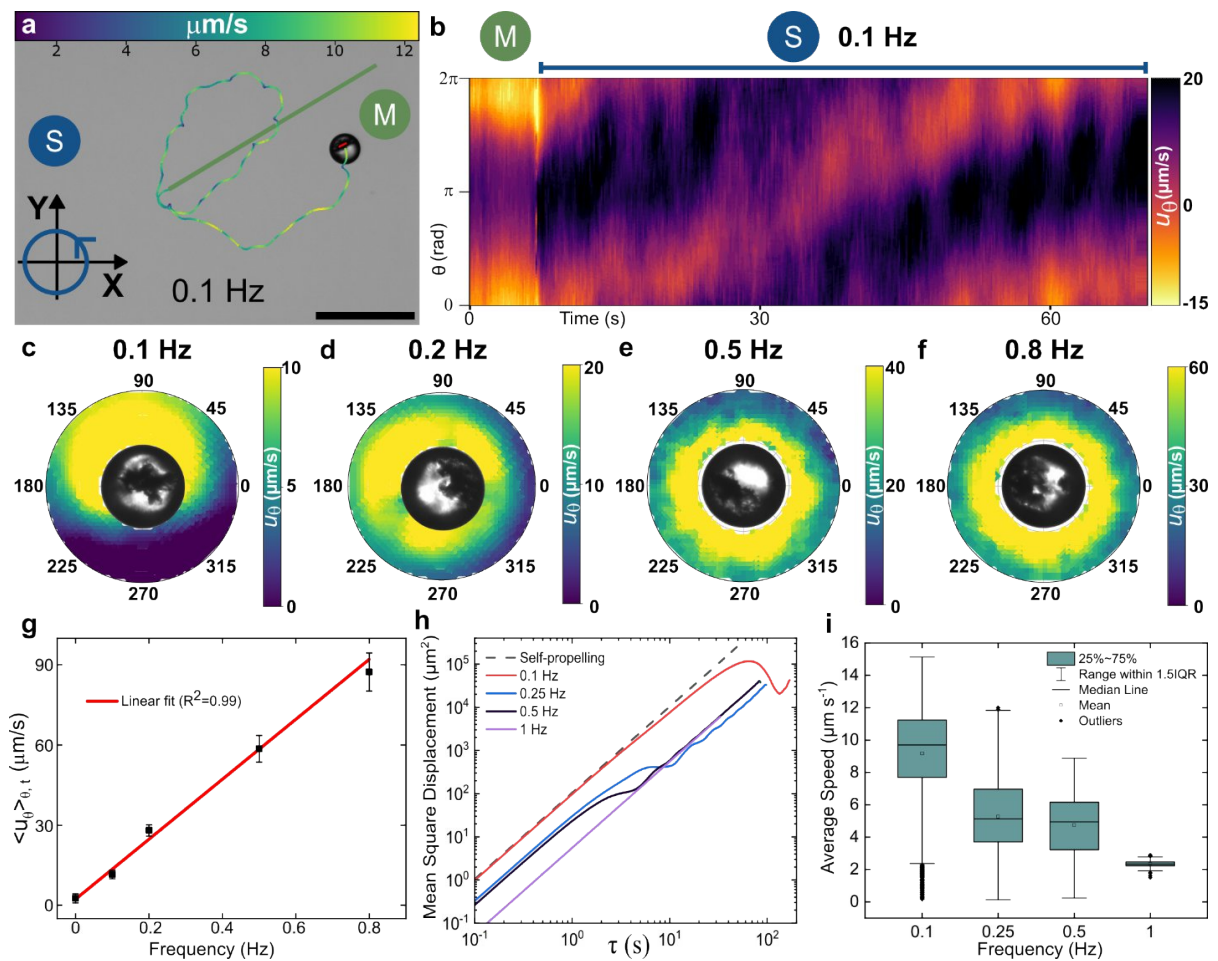

**Fig. 2. Curling motion of inherently achiral droplets.** (**a**) Switching from straight-line motion (M) to curling motion (S) using in-plane rotating fields (10 mT) at 0.1 Hz. Color bar represents the instantaneous speed in μm s$^{-1}$ and scale bar is 200 μm. (**b**) Kymograph showing values of $u_\theta$ (color) at a distance of 20 μm from droplet interface when switching from straight-line to the curling motion at 0.1 Hz. The two bands switch from being-parallel to x-axis (indicating a straight-line propulsion) to bands with positive slope, indicating counter-clockwise rotation of the dipolar symmetry axis of the Marangoni flows. (**c-f**) Representative plots (experimental snapshot of the droplet at the center) showing that the profile of $u_\theta$ becomes more uniform at higher frequencies. (**g**) Plot of $u_\theta$ averaged over angle and time ($<u_\theta>_{\theta,t}$) vs rotation frequency of the magnetic field. The values linearly increase with the frequency of the magnetic field because the strength of rotational flows linearly increases with rotation frequency. Error bars indicate standard deviation over time. (**h**) Mean square displacement data for self-propelling (straight-line) and curling droplets at different applied field frequencies. (**i**) Average instantaneous speeds of droplets during the curling motion. Error bars indicate the extent of oscillations (Supplementary Fig. 15 for more details).

Perturbing the Marangoni flows in droplets did not hinder their inherent properties like interactions with their trails or straight line-propulsion, but rather complemented their behaviors with curling motion and magnetic steering. Now we question about the influence of such flow perturbations on the chemotactic interactions between active droplets, as such chemotactic interactions lead to intriguing emergent collective behaviors of active droplets[5,9,10]. We begin with exploring droplet-droplet interactions of alike droplets (8CB) where both repulsive and attractive interactions co-exist. Repulsive interactions arise due to the chemo-repulsive clouds around the droplet and the solubilized droplet trails[13,14,21], while a fine hydrodynamic balance in the alignment of the Marangoni flows results in an attractive interaction leading to the formation of a propelling hydrodynamic pair/line[8,10]. To explore stability of this hydrodynamic pair, we steer two droplets until they reach a converging path and start to propel as a hydrodynamic pair perpendicular to the applied field (Fig. 3a). Afterwards, we change the direction of static fields to control the trajectory of the pair. Since, the hydrodynamic pair is a result of the temporary alignment of the Marangoni flows, we notice that the inter-droplet distance can increase after consecutive steering as the flows misalign and chemo-repulsive interactions dominate (Fig. 3b). Followed by a reduction in the distance once the flows align again, the hydrodynamic pair can be precisely controlled to propel over longer trajectories (Supplementary Video 5). Building on this finding, we further utilize their inherent tendency to align to steer a collective of many droplets (Fig. 3c). We notice that at higher number densities these droplets form longer chains that propel perpendicular to the applied field (Fig. 3c). Concurrently, these chains adapt to external magnetic field's direction, as well as break after consecutive steering and form new chains.

To probe further into externally influencing the hydrodynamic interactions between droplets, we investigate static collective of 5CB (4-Cyano-4'-pentylbiphenyl) droplets[8,22]. These droplet collectives remain static due to the stabilization of the flow fields around them and geometric constraints. Like bare 5CB droplets, the magnetic cluster doped 5CB droplets also form static droplet collectives (Fig. 3d I, Supplementary Video 6). Under static fields, the droplets reorient while keeping the collective state intact (Fig. 3d II). Furthermore, the collective translates while adapting to the magnetic cluster's new orientation providing control over both orientation and movement of the droplet collective (Fig. 3d III-IV). We then study the effect of rotating magnetic fields and observe that entire collective rotates about its axis as a single body at 0.2 Hz (Fig. 3e). Upon increasing the frequency to 0.5 Hz, we observe that collective splits and individual droplets curl away from each other, likely due to destabilization of hydrodynamic fields by the rotational flows (Fig. 3f). Once the rotational flows cease, droplets try to come closer to each other and reestablish the collective state, however, this attraction remains limited by the distance between them (Fig. 3g). Thus, we utilize such rotational flows to split existing collective states and disperse individual droplets to establish new collective states at higher droplet densities (Fig. 3h).

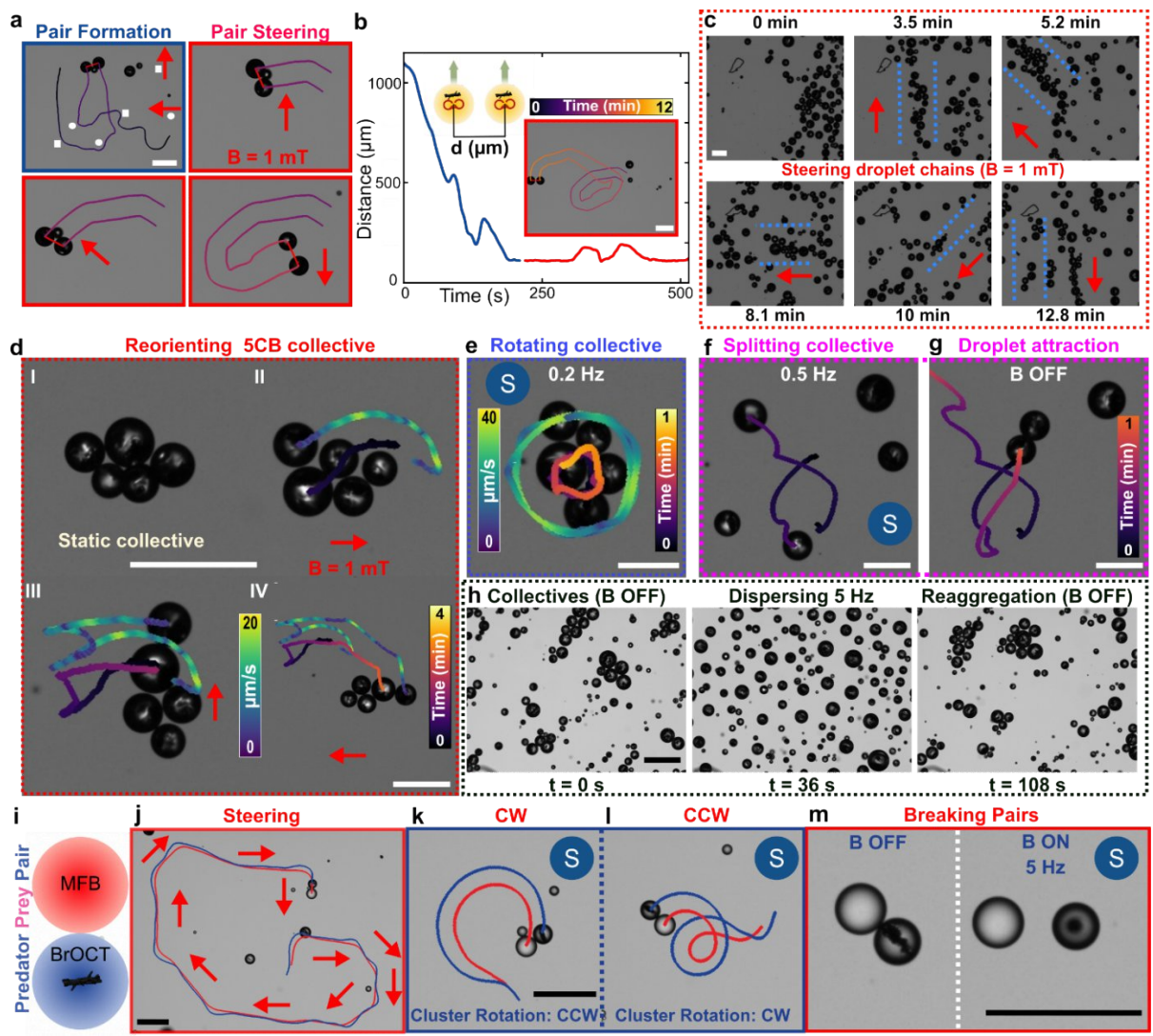

**Fig. 3. Controlling collective behaviors of various emulsion systems in synergy with inter-droplet interactions.** (**a**) Formation and steering of a hydrodynamically stabilized pair of 8CB droplets. (**b**) Plot showing the distance between the droplets in the pair during the steering. Inset: snapshot with trajectories (color bar represents time) of the two droplets during steering. (**c**) Steering of hydrodynamically stabilized chains of 8CB droplets. The chains grow along the applied field direction. Blue dashed lines are guide to the eye. (**d**) Reorientating a static collective of 5CB (4-Cyano-4'-pentylbiphenyl) droplets. (**e**) Rotating magnetic field at 0.2 Hz causes continuous rotation of 5CB droplet cluster. In (d and e), the trajectory of the outermost droplet is color-coded with its instantaneous speed and that of biggest droplet is color-coded with time. (**f**) Rotating magnetic fields at 0.5 Hz causes the collective to split and (**g**) switching off the magnetic field allows the droplets to attract and reaggregate. The trajectories in (f and g) are color-coded in time. (**h**) Dispersing 5CB droplet collectives using a rotating magnetic field at 5 Hz. The droplets slowly reaggregate on switching off the magnetic field. (**i**) Schematic of a predator-prey pair formed from a magnetic cluster-doped BrOCT (1-bromooctane) droplet and a bare MFB (methoxyperfluorobutane) droplet. (**j**) Consecutive steering of the predator-prey pair. (**k**) Clockwise curling motion of a pair under counterclockwise rotating magnetic field (1 Hz) and vice-versa (**l**). (**m**) The predator-prey pair (left) breaks under a rotating magnetic field at 5 Hz

(right). Red arrows show the direction of static magnetic field. Scale bar represents 300 $\mu$m in (h) and 200 $\mu$m in the rest.

The final test of our control strategy lies with emulsion systems where chemotactic behaviors do not remain limited to hydrodynamic interactions but in combination with material exchange enable non-reciprocal interactions between droplets[9]. 1-bromooctane (BrOCT, predator) droplets chase methoxyperfluorobutane (MFB, prey) droplets and propel as predator-prey pairs (Fig. 3i). We dope magnetic cluster to BrOCT droplets as the oil chemistry allows for encapsulation of FePt magnetic particles. In 0.5 wt% Triton X 100 (200 times CMC), magnetic cluster doped BrOCT droplets only propel when MFB droplets are in vicinity, which is consistent with the behavior of bare BrOCT droplets[9,23]. In addition, the droplets form a chasing pair as they reach closer to each other and continue to propel in this configuration. As we perturb the magnetic cluster with static fields, BrOCT droplet slightly rotates at the interface of the MFB droplet influencing the pair to adapt to a new direction. This is likely as oil-exchange between the two droplets occurs from the rear of the prey (MFB) and the pair adapts accordingly to keep BrOCT droplets at the rear. In contrast to previous emulsion systems, the chasing pair propels along the field direction, likely due to different internal flow profiles. Nonetheless, multiple steering leads to a precise control over the chasing pair's trajectory (Fig. 3j and Supplementary Video 7). We further exploit the pair's ability to maintain their chasing configuration and continuously rotate BrOCT droplets on the interface of MFB droplets by using rotating magnetic fields (Fig. 3k and l). The pair adapts to curling motion while the chirality of this curling motion is opposite to the magnetic cluster's direction of rotation. This finding is also coherent with the pairs attempt to keep BrOCT droplets at the rear and maintain the non-reciprocal oil exchange. Furthermore, we demonstrate an on-demand breakage of the chasing interactions by utilizing strong rotational flows (Fig. 3m). Stronger rotational flows dominate the Marangoni flows and diminish non-reciprocal oil-exchange that keeps the chasing pair configuration intact.

**Discussion and Conclusion:**

Overall, we introduced a synthetic design to reach a combination of autonomous and externally driven behaviors in emulsion droplets, including spontaneous self-propulsion, magnetic steering, curling, and perturbing collective behaviors. In contrast to solid active matter where external perturbation leads to its synchronous rotation[24], our strategy relies on reorientation of the Marangoni flows by utilizing the mechanical coupling between these flows and the cluster. While such reorientation provides control over the motion of the droplet, the inherent chemotactic behaviors of these droplets remain unaffected. Thus, providing a tool to control and perturb their emergent behaviors. The generality of this strategy also removes barriers of specific chemistries and allow for probing into dynamics of various emulsion systems.

Despite the richness and adaptability of our strategy, a lot remains to be explored. For example, our results remain limited to the regime where droplet propulsion remains steady and the extension of our perturbation strategy to domains where droplet propulsion becomes unsteady (i.e. when flows become chaotic) requires further exploration (for example, propulsion at high Péclet numbers[18]). Additionally, it will be interesting to investigate solid-fluid interactions at fluidic interfaces with non-trivial geometries (for instance, liquid-fibers[25] and dynamic interfaces[26]). Simultaneously, our results open intriguing questions on understanding the behavior of fluidic flows when subjected to complex mechanical perturbations.

As a control strategy, we utilize uniform global magnetic fields, where all droplets experience the same perturbation. In complex confined spaces, each droplet experiences a different local environment and may require an independent control. The utilization of non-uniform magnetic field[27–30] could prove efficient for controlling multiple droplets independently in such scenarios.

Thus far, we demonstrated how various droplet collectives can be perturbed using our strategy. However, more sophisticated magnetic fields[31] may yield further insights into the forces governing collective dynamics, and open room for unprecedented emergent behaviors[32,33]. Our strategy may supplement the use of active emulsions as an experimental model to further develop fundamental insights on the locomotion of single-celled microorganisms[34,35] near boundaries[36–38] and in complex fluids[39–41]. Finally, combined autonomous and externally driven behavior of droplets is a steppingstone to realize intelligent and adaptive micromachines. Further studies on translation of these control strategies to reaction-driven biocompatible emulsions may be useful in biomedical technologies[42].

## Materials and Methods

### Materials

We used following chemicals as received without any further purification or modification: 4-octyl-4'-cyanobiphenyl (8CB, Synthon, 99.5 %), 1-Bromohexadecane (Sigma-Aldrich, 97%), 4-Heptyloxybenzaldehyde (TCI Chemicals, 98%), 4-Cyano-4'-n-Pentylbiphenyl (5CB, Thermo Scientific Chemicals 99%), 1-Bromooctane (Sigma-Aldrich, 99%), Methoxyperfluorobutane (Sigma-Aldrich, 99%, mixture of n- and iso-butyl isomers), Triton™ X-100 (Sigma-Aldrich) Tetradecyltrimethylammonium bromide (TTAB, Sigma-Aldrich), Hexadecyltrimethylammoniumbromide (CTAB, Sigma-Aldrich), Hexadecyl-trimethylammoniumchloride (CTAC, Sigma-Aldrich) Platinum(II) acetylacetonate (Pt(acac)2, Sigma-Aldrich, ≥99%), Iron(III) acetylacetonate (Fe(acac)3, Sigma-Aldrich, ≥99%), Oleic acid (Sigma-Aldrich, 90%), Oleylamine (Thermo Scientific Chemicals, 80-90%), Polystyrene particles (Microparticles GmbH: 0.628 µm, PS-R-0.65).

### Synthesis of FePt nanoparticles

We synthesize $L_{10}$ phase FePt nanoparticles as reported by Yu, Y. et al, with slight modifications[43]. Briefly, we mix 0.2 mmol of platinum (II) acetylacetonate, 0.2 mmol of iron (III) acetylacetonate and 0.1 mmol of cetyltrimethylammonium chloride in a round-bottom flask with 0.5 ml Oleic acid (OA) and 10 ml of Oleylamine. Afterwards, we purge the mixture with argon gas for 1 hour to provide inert atmosphere. The mixture is heated up to 340°C under reflux condition and in the presence of Argon. These reaction conditions are maintained for 3 hours followed by subsequent cooling to room temperature by removing the heating mantle. FePt nanoparticles are extracted from the reaction mixture by washing with hexane: ethanol (1:2 V/V) mixture for five times using centrifugation. FePt particles are stored by dispersing in chloroform at room temperature conditions.

### 8CB (4-Cyano-4'-octylbiphenyl) droplet production

We further purify as synthesized FePt nanoparticles using magnetic separation and by washing them with chloroform. Then, 20 mg of FePt particles are obtained after drying chloroform in a glass vial. These particles are dispersed in 1 ml of 8CB oil using sonication at 45°C. 10 µl of this mixture is placed in 1 ml of 1 wt% TTAB solution pre-heated at 45°C. Droplets are produced using shake emulsification method. The disadvantage of this method is polydispersity in size of droplets and magnetic particle distribution. To overcome this, we first separate magnetically active droplets using a permanent magnet and pipette approximately 20 µl of the emulsion solution from the side close to the permanent magnet and transfer them to another vial containing 1 ml of 1 wt% TTAB. With the reduced concentration of droplets, we gently swirl the vial to disperse droplets in the entire volume and wait to allow time for size-based distribution. Afterwards, we pipette 3 µl from near the bottom of the vial to extract only a few of the droplets with desired diameter. The number of droplets per sample is reduced to less than ten to reduce the droplet-droplet interactions and allow us to focus on a single droplet. We use a single batch of the droplets for 2 days of experiments and after that, we prepare a new batch of droplets for new experiments.

**HBA (4-Heptyloxybenzaldehyde) droplet production**
We disperse purified 20 mg FePt nanoparticles in 1 ml of HBA oil via sonication at room temperature. 10 µl of this mixture is placed in 1 ml of 1 mM CTAB solution at room temperature. Droplets are produced using shake emulsification method. Later, magnetically active droplets are separated using a permanent magnet and stored in 1 mM CTAB solution before carrying out the experiments. Each batch is prepared freshly before the experiments.

**1-bromohexadecane droplet production**
We disperse 20 mg of purified FePt nanoparticles in 1 ml of 1-bromohexadecane oil via sonication at room temperature. 10 µl of this mixture is placed in 1 ml of 0.05 wt% Triton X-100 solution at room temperature. Droplets are produced using shake emulsification method. Later, magnetically active droplets are separated using a permanent magnet and stored in 0.05 wt% Triton X-100 solution before carrying out the experiments. Each batch is prepared freshly before the experiments.

**5CB (4-Cyano-4'-pentylbiphenyl) droplet production**
We disperse 20 mg of purified FePt nanoparticles in 1 ml of 5CB oil via sonication at room temperature. 10 µl of this mixture is placed in 1 ml of 0.1 wt% TTAB solution at room temperature. Droplets are produced using shake emulsification method. Next, magnetically active droplets are separated using a permanent magnet and stored in 0.1 wt% TTAB solution before carrying out the experiments. Each batch is prepared freshly before the experiments.

**Experimental protocols**
We carry out self-propulsion experiments in a glass chamber (length (l) and width (w) = 12 mm) of height (h) 160 µm, to limit the movement in z-axis.

**1) Cluster motion in 8CB droplets**
8CB droplets are mixed with 10 wt% TTAB solution and introduced into our experimental glass chamber. We prepare the samples at room temperature while the droplets are in smectic phase (8CB: Smectic to Nematic transition temperature, $T_{S-N}$ = 33.5°C) and demonstrate no self-propulsion. Afterwards, the chamber is sealed and transferred to an optical microscope (Zeiss; Axio Imager 2) equipped with a temperature chamber and a 5-Axis electromagnetic coil setup. We heat up the experimental chamber till 37°C (8CB: Nematic window, $T_N$ = 33.5-38°C) to initiate self-propulsion in the nematic state. Under no magnetic field, droplets propel in random directions due to isotropic environmental conditions (homogeneous surfactant distribution). To observe the influence of cluster's orientation on the Marangoni flows, we align the cluster using 5 mT static field, while the droplet is in smectic state (not propelling). Afterwards, we heat-up the experimental chamber till 37°C and record the movement of the droplets and the clusters.

## 2) Magnetic steering and curling

We first introduce 8CB droplets dispersed in 1 wt% TTAB solution into the glass chamber and fill the chamber with 20 wt% TTAB from the other end, to provide isotropic environment to the droplets. We prepare the samples at room temperature while the droplets are in smectic phase (8CB: Smectic to Nematic transition temperature, $T_{S-N}$ = 33.5°C) and demonstrate no self-propulsion. Afterwards, the chamber is sealed and transferred to an optical microscope (Zeiss; Axio Imager 2) equipped with a temperature chamber and a 5-Axis electromagnetic coil setup. Temperature is maintained at 37°C (8CB: Nematic window, $T_N$ = 33.5-38°C) to initiate self-propulsion in the nematic state. We apply static fields (in-plane) for steering and in-plane rotating fields for curling motion using electromagnetic coils. In all experiments with 8CB droplets, we use TTAB as stabilizer for the emulsions and initiator for self-propulsion (at higher concentrations: above 5 wt %).

HBA, 1-bromohexdecane and 5CB droplets are also studied in a similar experimental glass chamber (l & w = 12 mm, h = 160 μm). Experiments involving these three oils are carried out at room temperature. The only difference arises in the choice of surfactant for each of these droplets. Briefly, a small amount of HBA droplets dispersed in 1 mM CTAB solution are introduced into the glass chamber and the chamber is then filled with 10 mM CTAB solution. In case of 1-bromohexdecane droplets, initially droplets are introduced in the chamber while being dispersed in 0.05 wt% Triton X-100. Next, the chamber is filled with 0.5 wt% Triton X-100 solution. 5CB droplets are initially dispersed in 0.1 wt% TTAB solution and then the chamber is filled with 5 wt% TTAB solution. In all cases, droplets immediately begin to propel once they come into contact with higher surfactant concentrations. The chambers are then sealed and studied under the microscope equipped with the electromagnetic coil setup. We apply static fields of strength 1 mT to control the direction of self-propulsion of each of these droplets.

## 3) Steering 8CB pairs and lines

To form pairs, we increase the concentration of the droplets slightly inside the glass experimental chamber (l & w = 12 mm, h = 160 μm) to the extent that droplet-droplet interactions dominate. To form lines, we increase the concentration of droplets significantly and reduce the dimension of the chamber (circular chamber with diameter = 2.5 mm, h = 160 μm). We carry out these experiments at 37°C in 20 wt % TTAB solution. We apply static field (1 mT) to control individual, pairs and lines of droplets.

## 4) 5CB collectives

To form the 5CB droplet collectives, we followed the protocols as previously reported[8,22] Briefly, we prepare an experimental chamber composed of cylindrical PDMS spacer of diameter 2.5 mm and height 5 mm mounted on a glass substrate. The chamber is prefilled with 10 wt% TTAB solution. Concentrated solution of 5CB droplets dispersed in 0.1 wt% TTAB solution are introduced inside this experimental chamber. After allowing time for the droplets to settle at the bottom, we then seal the chamber using a glass cover slip. The sample is then placed under the microscope for further observation and application of magnetic fields. We apply static fields (1 mT) to reorient the established collective states, rotational fields (5 mT in strength) at 0.2 Hz to rotate the collective state and at 0.5 Hz to split the collective states. Increasing the concentration of the droplets, increases the observable collective states and we utilize strong rotational flows (5 HZ) to split collective states, disperse droplets and form new collective states.

**5) Predator-prey pair (steering, curling, and splitting)**

We dope FePt nanoparticle cluster to 1-bromoocatane droplets (predator droplets). Briefly, we disperse FePt nanoparticles inside 1-bromoocatane droplets at a concentration of 20 mg/ml via sonication. 10 µl of this mixture is placed in 1 ml of 0.5 wt% Triton X-100 solution at room temperature. Droplets are produced using shake emulsification method. Later on, magnetically active droplets are separated using a permanent magnet and transferred to a fresh solution of 0.5 wt% Triton X-100. Immediately afterwards, we introduce these droplets into our experimental glass chamber (l & w = 12 mm, h = 160 µm). From the opposite side, we introduce relatively similar concentration of Methoxyperfluorobutane (MFB, prey) droplets. Briefly, MFB droplets are produced by dispensing 10 µl of MFB oil in 0.5 wt% Triton X-100. The mixture is subjected to vigorous shaking to produce MFB droplets. The remaining of the chamber is filled with 0.5 wt% Triton X-100 solution. In our experimental conditions, 1-bromoocatane droplets migrate towards the region containing majority of the MFB droplets. Once a predator-prey pair is formed, we apply static fields (1 mT) to steer the trajectory of the droplets, rotating fields at 1 Hz to induce curling motion and split the pairs by applying strong rotational flows at 5 Hz.

**Droplet tracking and analysis**

Experimental videos are analyzed using a custom Python script and openCV libraries. Particle speeds and trajectories are extracted from the detected particle positions in each frame.

**Particle image velocimetry (PIV) analysis**

Experiments videos are recorded for droplets dispersed in 10 wt% TTAB solutions containing ~628 nm Polystyrene tracer particles (10 times diluted from stock solution of 5wt%) and the recorded video frames are used for the PIV analysis. PIV analysis is performed using Dynamic studio software (version 6.1). The droplet is masked in the images using phase boundary detection and Adaptive PIV analysis is applied with minimum and maximum IA size being16 x 16 and 8 x 8, respectively and Grid step size being 8 x 8. Then, a 5 x 5 average filter is applied to the results of the adaptive PIV, followed by vector resampling (step size = 5). The final results are overlaid with the experimental images to visualize flows around a droplet in the laboratory frame. The laboratory frame results are further analyzed using a custom Python script to calculate tangential flows ($u_\theta$) in the translating droplet frame. The x-axis of the translating frame is always aligned to the x-axis of the laboratory frame. The framerate for experimental videos used for PIV analysis is 6.5 frames per second (fps) for steering and 0.1 Hz and 0.2 Hz curling experiments and 65 fps for 0.5 Hz and higher frequency curling experiments.

**Acknowledgments:** We thank G. Richter and M. Yunusa for discussions regarding liquid crystalline materials, A. Shiva for help in nanoparticle magnetic property and TEM characterization, A. Pal for discussions regarding material characterization, and U. Bozuyuk, M. Han and S. Baltaci for discussions on optical microscope imaging.

**Funding:** This work is funded by the Max Planck Society. M. T. A. K. thanks the International Max Planck Research School for Intelligent Systems for financial support. M. Z. thanks the Alexander Von Humboldt Foundation. R. H. S. and M.S. thank the European Research Council (ERC) Advanced Grant (SoMMoR project, grant no.: 834531).

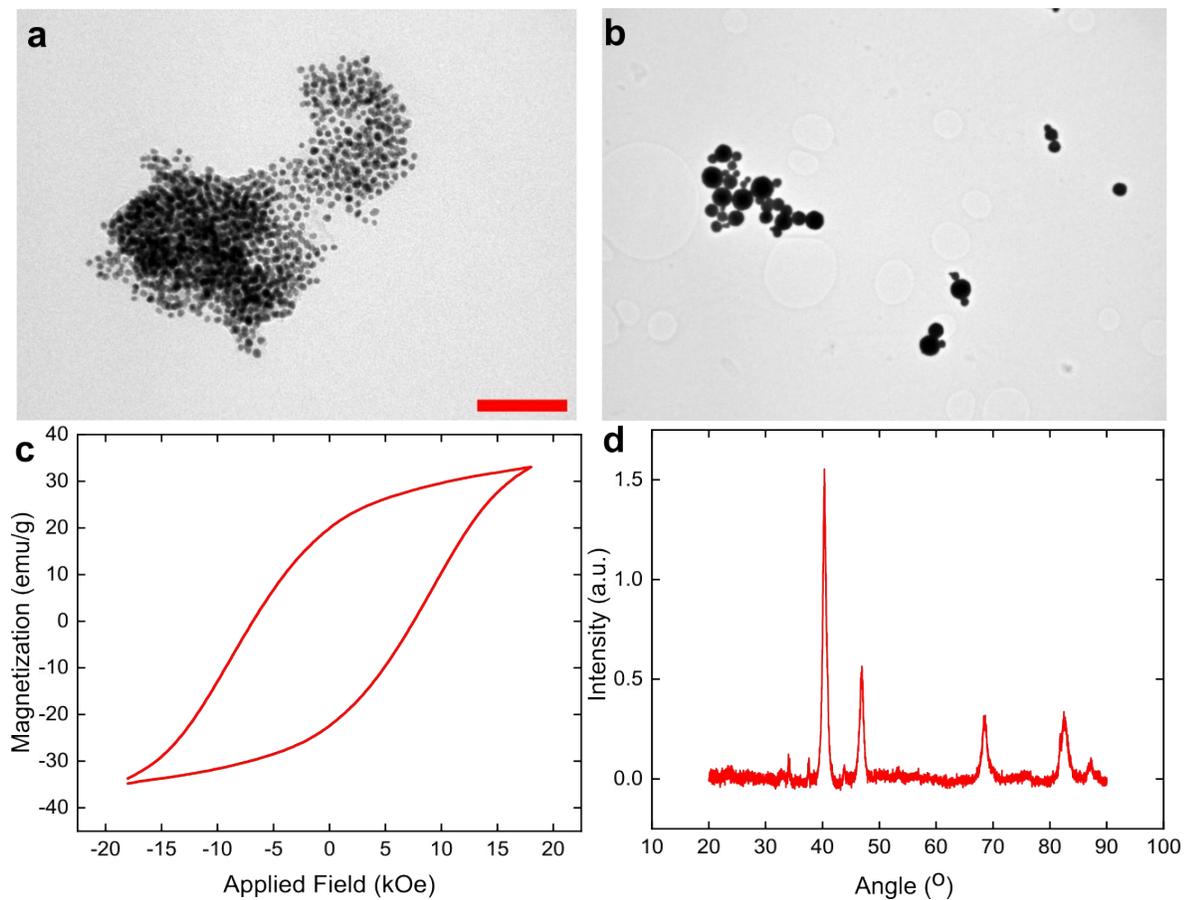

**Supplementary Fig. 1. Characterization of size, magnetic properties and crystalline structure of FePt particles.** (**a**, **b**) As-synthesized TEM images of FePt nanoparticles, showing presence of single 20 nm particles (**a**) and agglomerated particles (**b**). Scale bar represents 100 nm. (**c**) Magnetic properties of FePt nanoparticles, with coercive field of 718 mT. (**d**) XRD spectrum showing the $L_{10}$ ordering of FePt nanoparticles.

**Supplementary Note 1**

**1.1 Self-assembly of magnetic particles within static nematic droplet**

At low micellar concentrations, droplet remains static and possess topological point defect with +1 charge (radial hedgehog, (Supplementary Fig. 2) due to homeotrpoic anchoring of LC molecules by TTAB molecules adsorbed at LC-water interface. In LC media, foreign particles assemble at the point defect, due to lack of orientational ordering at the point defect. Similarly, the assembly of FePt particles originates at the point defect, with complex architecture pointing radially downwards as a cluster (Supplementary Fig. 2b and 2d). FePt appear as a cluster in the nematic environment, likely due to strong magnetic interactions between them and minimization of elastic interactions with surrounding LC molecules.

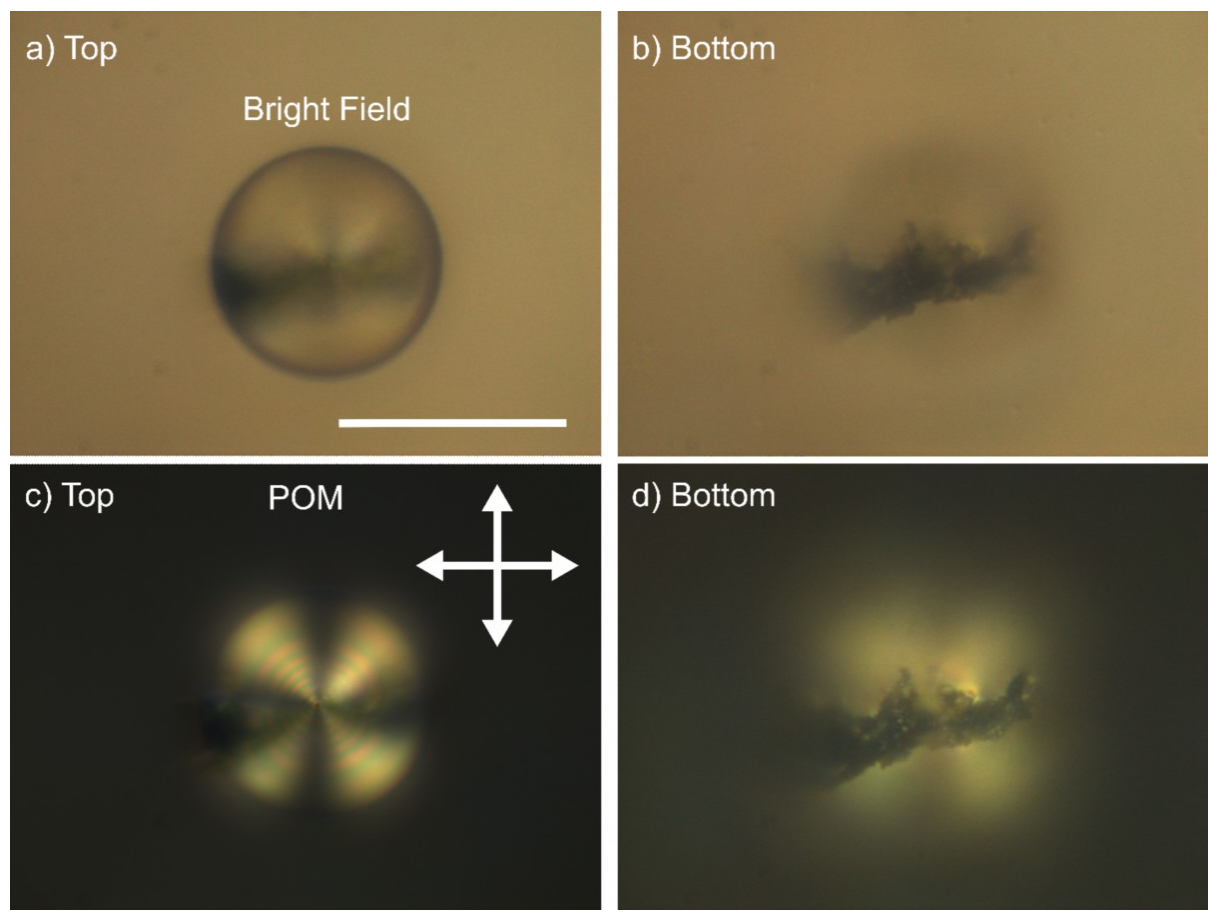

**Supplementary Fig. 2. Bright field and polarized optical microscope images (POM) of static LC droplet.** (**a**) Bright field image with focus on top of the droplet. (**b**) Bright field image with focus on bottom of the droplet. (**c**) POM image of droplet demonstrating topological point defect with +1 charge (focus on top of the droplet). (**d**) POM image with focus on bottom of the droplet. Scale bar represents 100 μm.

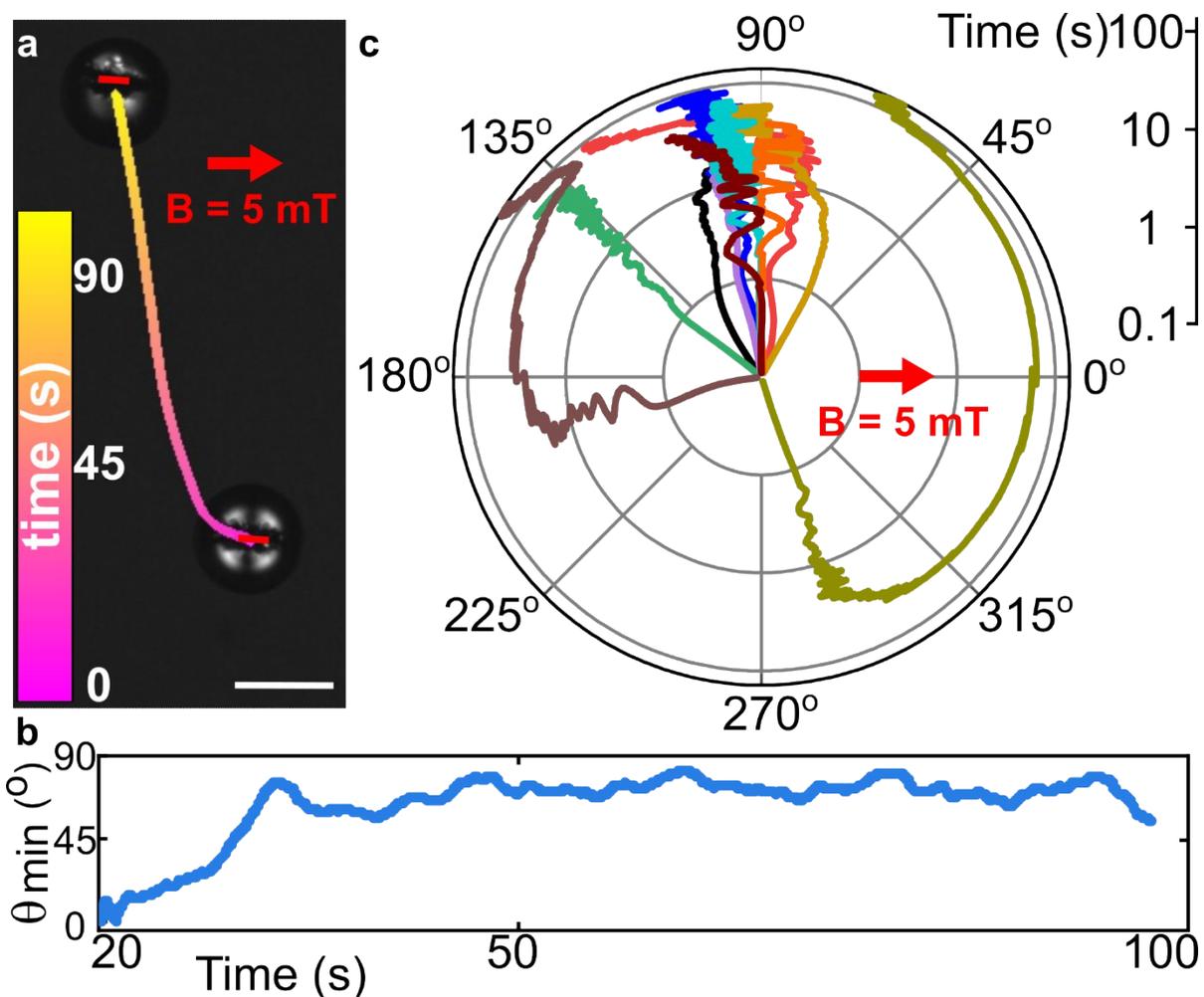

**Supplementary Fig. 3. Propulsion of droplets when magnetic cluster is pre-aligned before the Marangoni flows establish.** (**a**) Trajectory of an 8CB droplet just after transitioning to nematic state. The magnetic cluster was pre-aligned with a static magnetic field (5 mT) while the droplet was in the smectic state and before the Marangoni flows were established. Color bar represents time and scale bar represents 100 μm. (**b**) Evolution of the angle between the longitudinal axis of the cluster and the propulsion direction of the droplet ($\theta_{min}$) corresponding to the experiment shown in (a). $\theta_{min}$ gradually reaches 90°. Since the cluster's orientation is fixed with external magnetic field, the saturation of $\theta_{min}$ suggests that the Marangoni flows inside the droplet adapt to the cluster's orientation, causing the droplet to propel perpendicular to the applied magnetic field. (**c**) Evolution of the propulsion direction of 11 droplets (each indicated by a different color) over time, where 0° corresponds to the direction of the external magnetic field. The magnetic clusters in all the droplets were pre-aligned similarly to the droplet shown in (a).

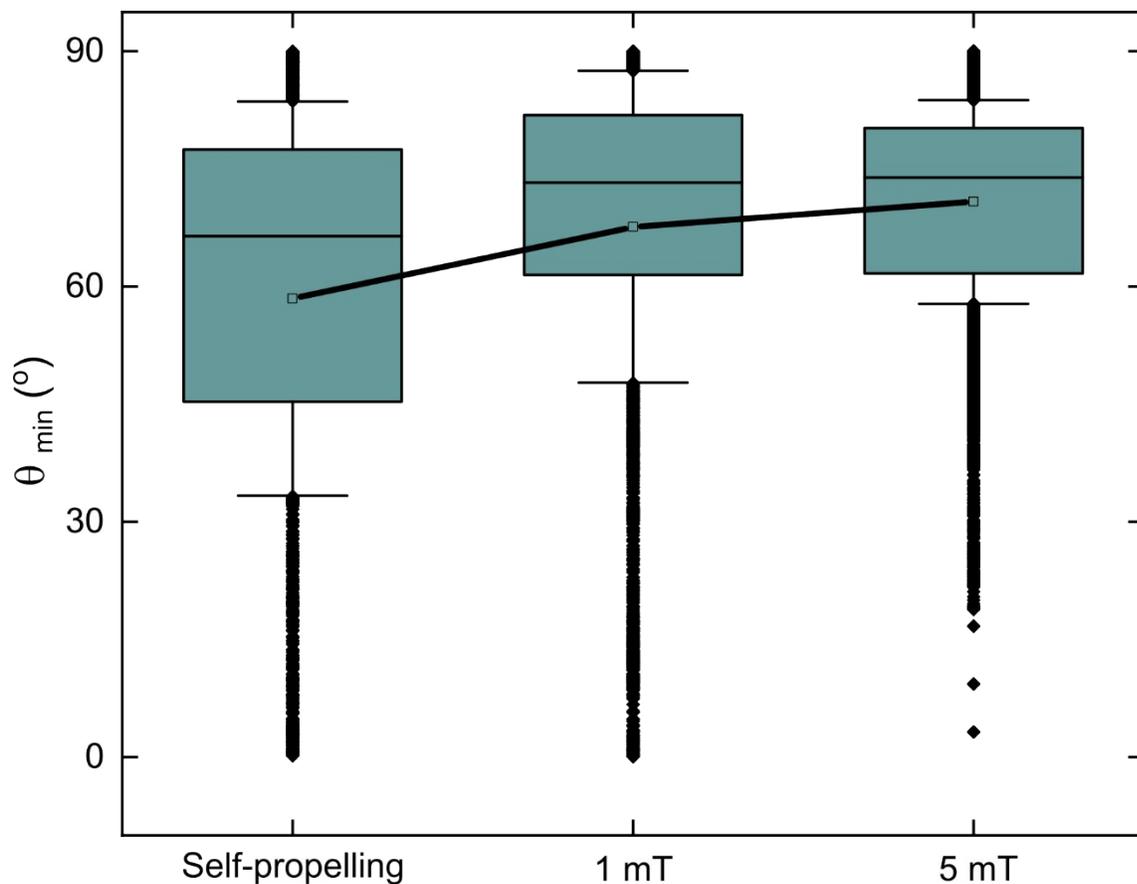

**Supplementary Fig. 4. Box plot of the angle between the longitudinal axis of the magnetic cluster and the propulsion direction of the droplet ($\theta_{min}$) for different values of static magnetic field.** The values are collected from trajectories of multiple droplets. The upper and lower bounds of the box represent 25 and 75 percentile and the error bars represent standard deviation over 9 droplets. $\theta_{min}$ is more bounded when external magnetic field is applied.

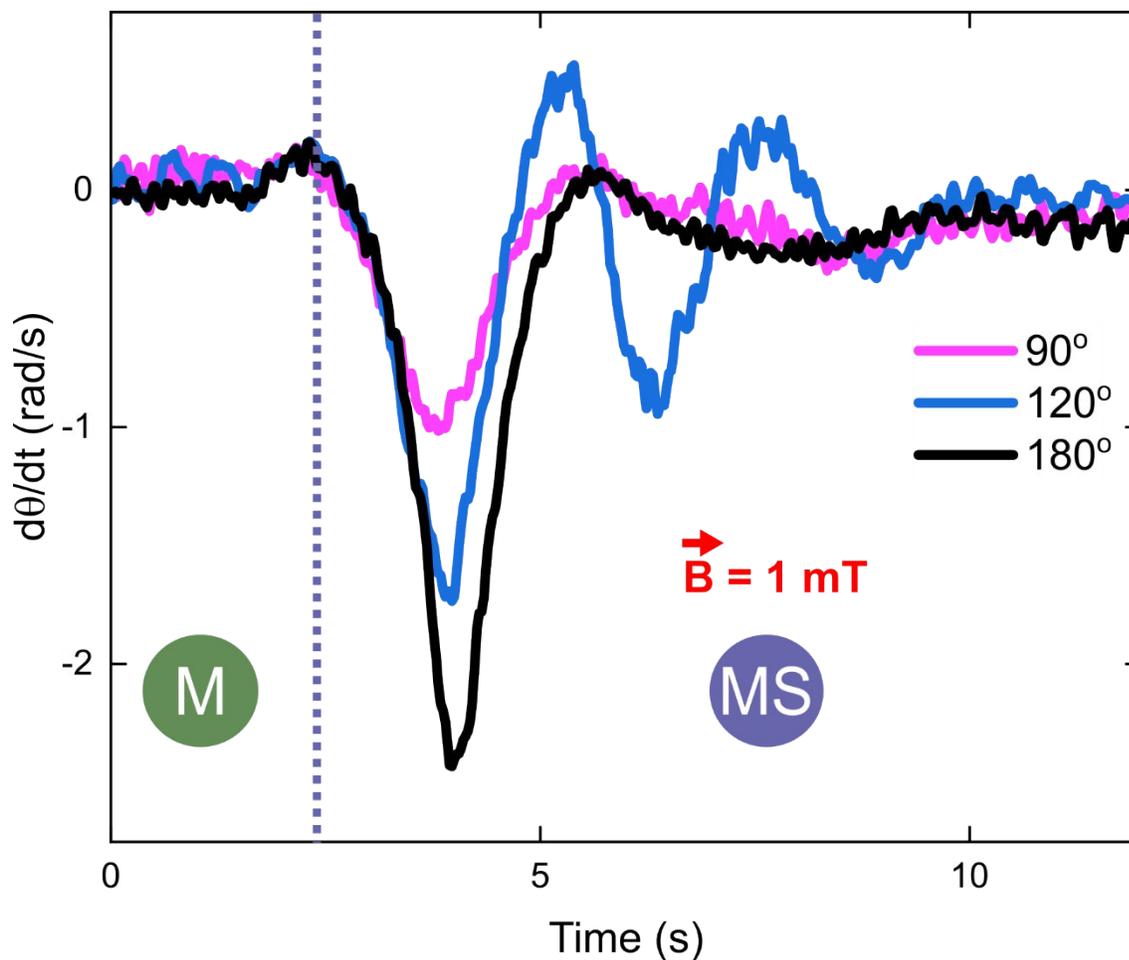

**Supplementary Fig. 5. The time derivative of the propulsion direction (d$\theta$/dt) of the droplet vs time.** The three plots correspond to cases when a droplet is steered by 90°, 120°, and 180°, respectively. d$\theta$/dt is close to zero in the beginning, corresponding to the straight-line propulsion of the droplet and it dips when the magnetic field is applied. Later, it slowly relaxes to 0. The time required for d$\theta$/dt to relax to zero varies with the extent of steering.

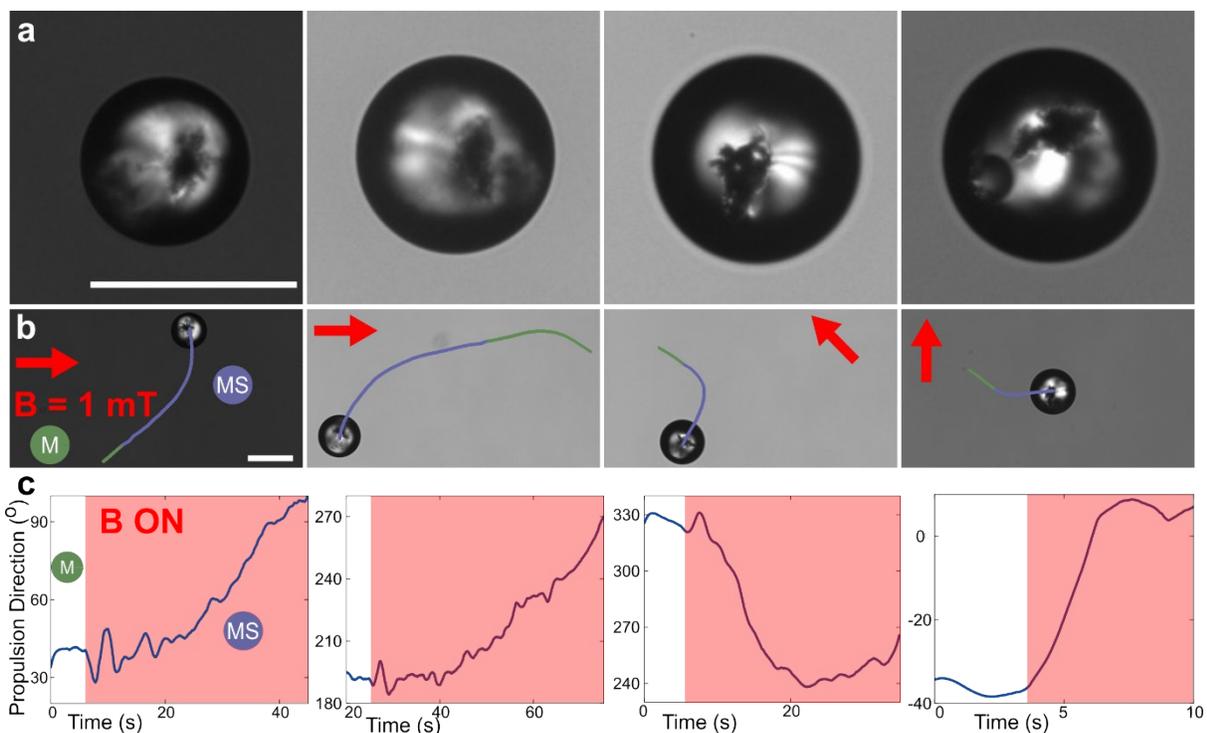

**Supplementary Fig. 6. Effect of cluster size on the magnetic steering of the droplets.** (**a**) Four droplets of similar diameter, with increasing cluster size from left to right. (**b**) Respective trajectories of the four droplets under the influence of static fields. Droplet undergoing self-propulsion (M) changes its direction upon reorienting the magnetic cluster (MS). Red arrows denote the direction of static magnetic field. (**c**) Evolution of the respective propulsion directions of the four droplets. A steady propulsion direction indicates that droplet is propelling in straight-line. Upon perturbing the magnetic cluster, propulsion direction changes. The time it takes for the propulsion direction to reach a new steady value is also a measure of steering time. Steering time decreases as the cluster size increases.

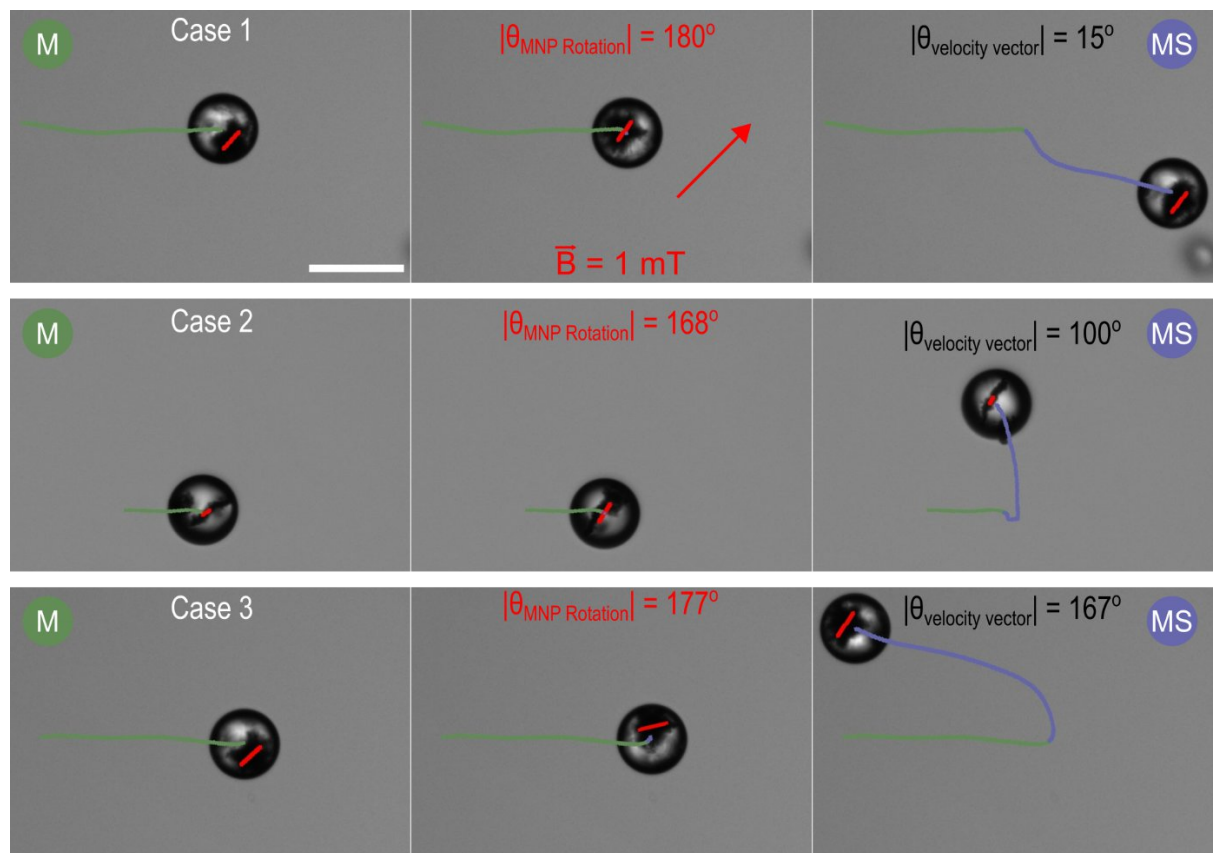

**Supplementary Fig. 7. Three cases for magnetic cluster rotations greater then 160º and resulting change in the velocity vector.** Scale bar represent 100 μm.

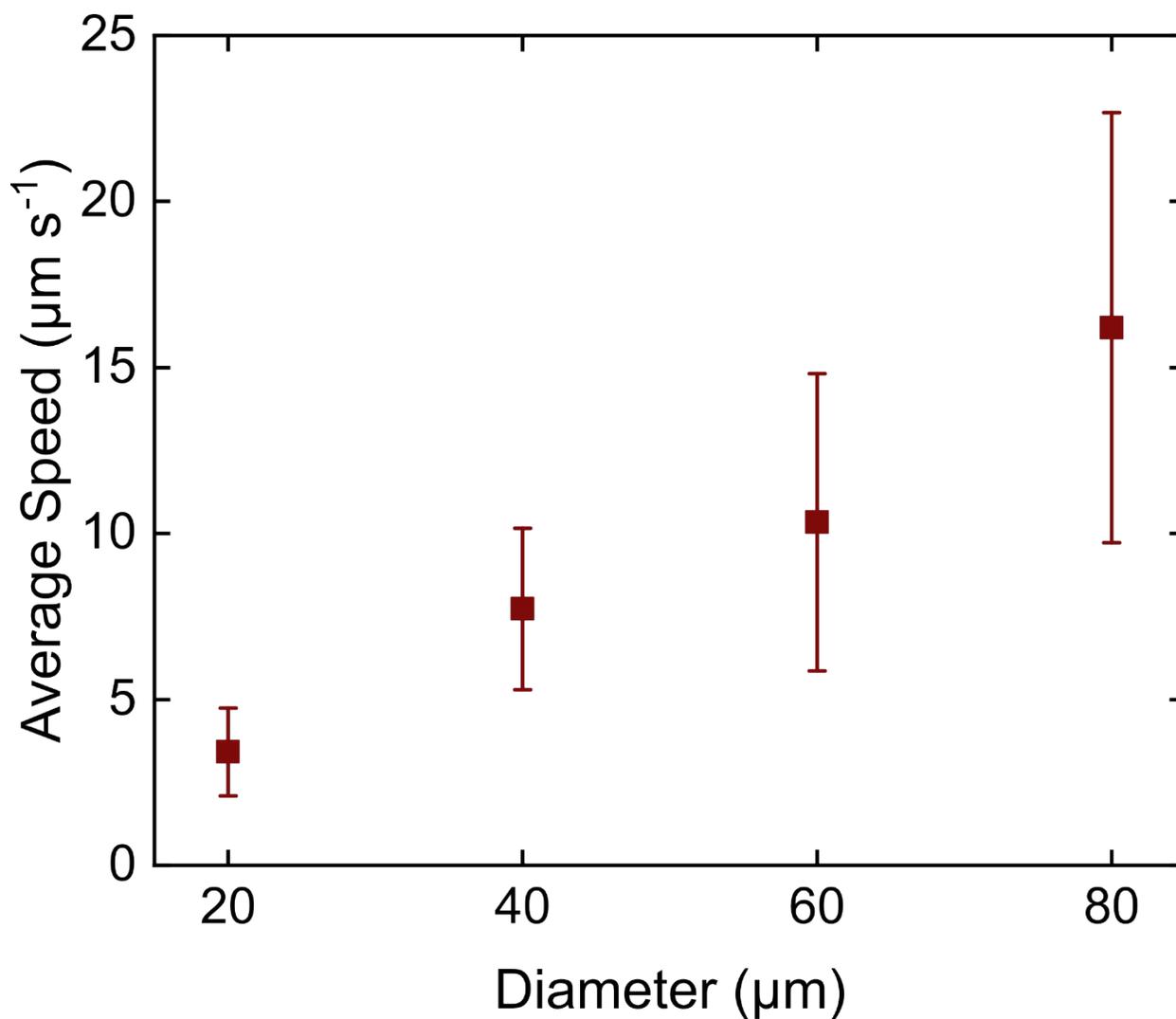

**Supplementary Fig. 8. Average speeds of self-propelling droplets at 20 wt% TTAB with respect to their diameter.** Error bars represent standard deviation over time.

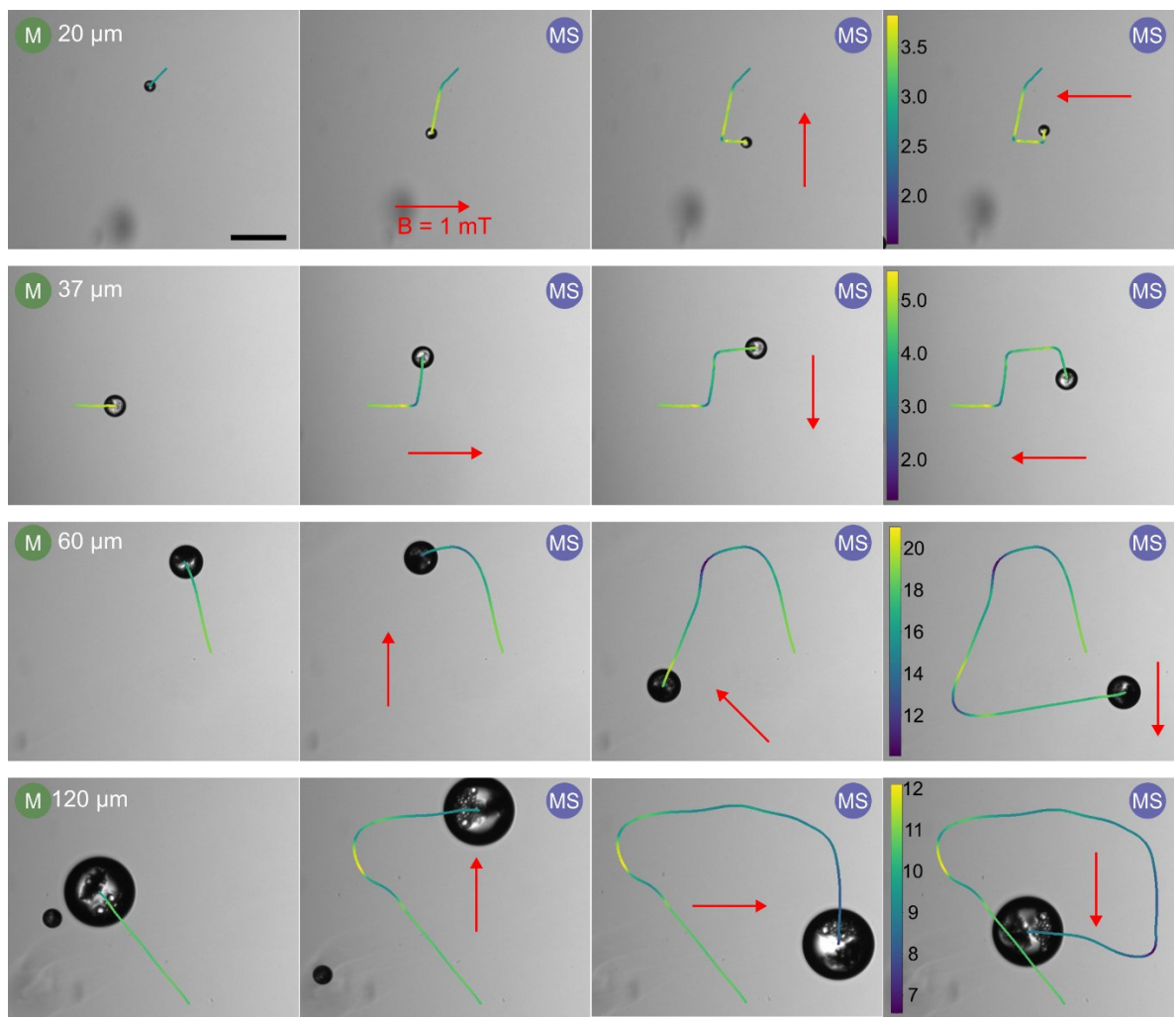

**Supplementary Fig. 9. Magnetic steering of self-propelling droplet of different diameters using static fields in x-y plane with field strength of 1 mT.** Experiments are performed in an infinity chamber of height 2 mm and surfactant concentration at 20 wt% TTAB. Trajectories of droplets are color coded with their instantaneous speed and color bar represents speeds in μm s$^{-1}$. Scale bar represent 100 μm.

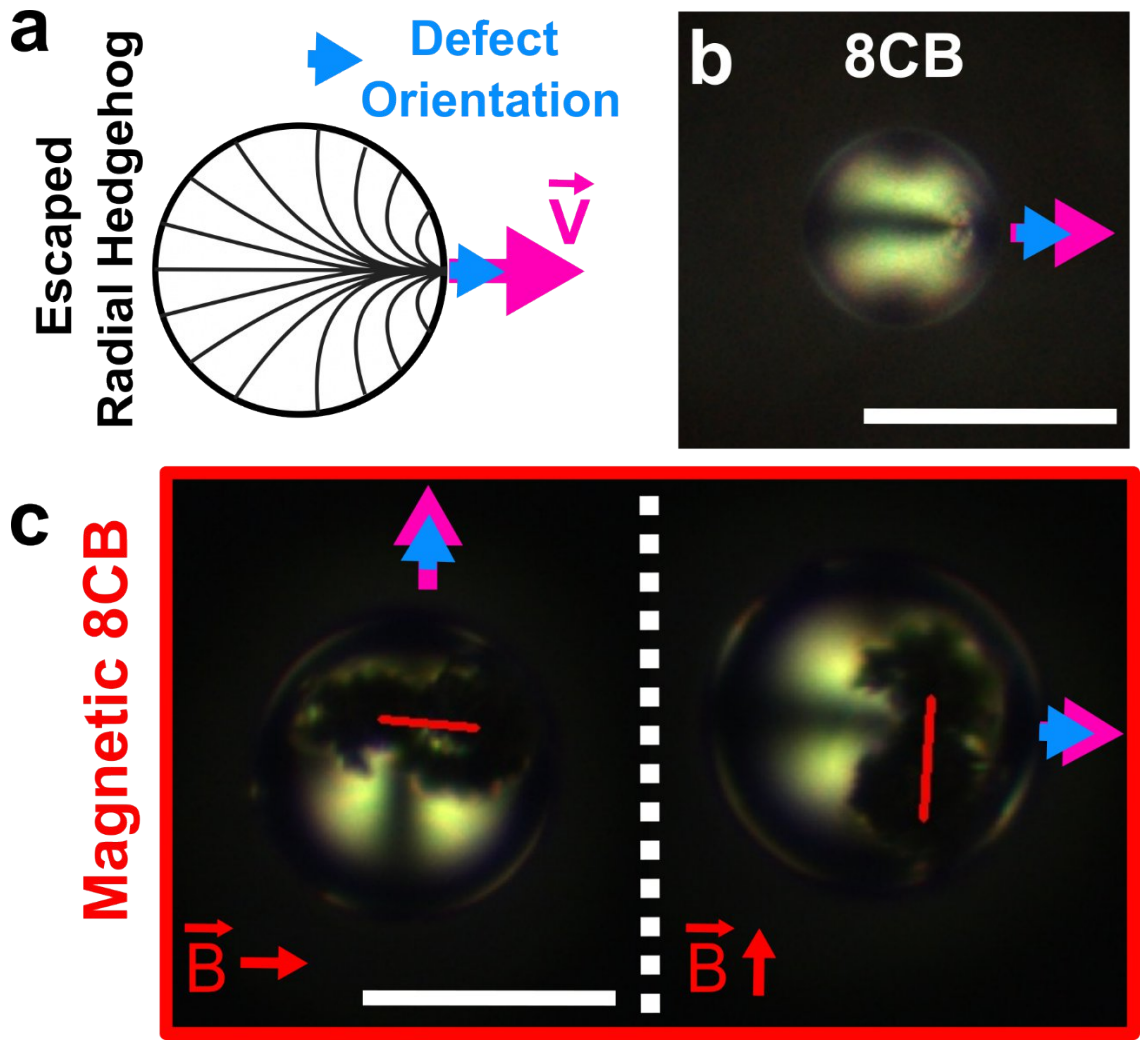

**Supplementary Fig. 10. Internal flow structure of the droplet during steering.** (**a**) Schematic showing that the point defect orients towards the propulsion direction (escaped radial hedgehog defect). (**b**) POM texture of a bare 8CB droplet during propulsion. (**c**) POM textures of a magnetic particle-doped 8CB droplet during steering, The defect orients along the direction of propulsion and perpendicular to the longitudinal axis of the cluster. The cyan and magenta arrows represent the defect orientation and the direction of propulsion of the droplet. Scale bars represent 100 $\mu$m.

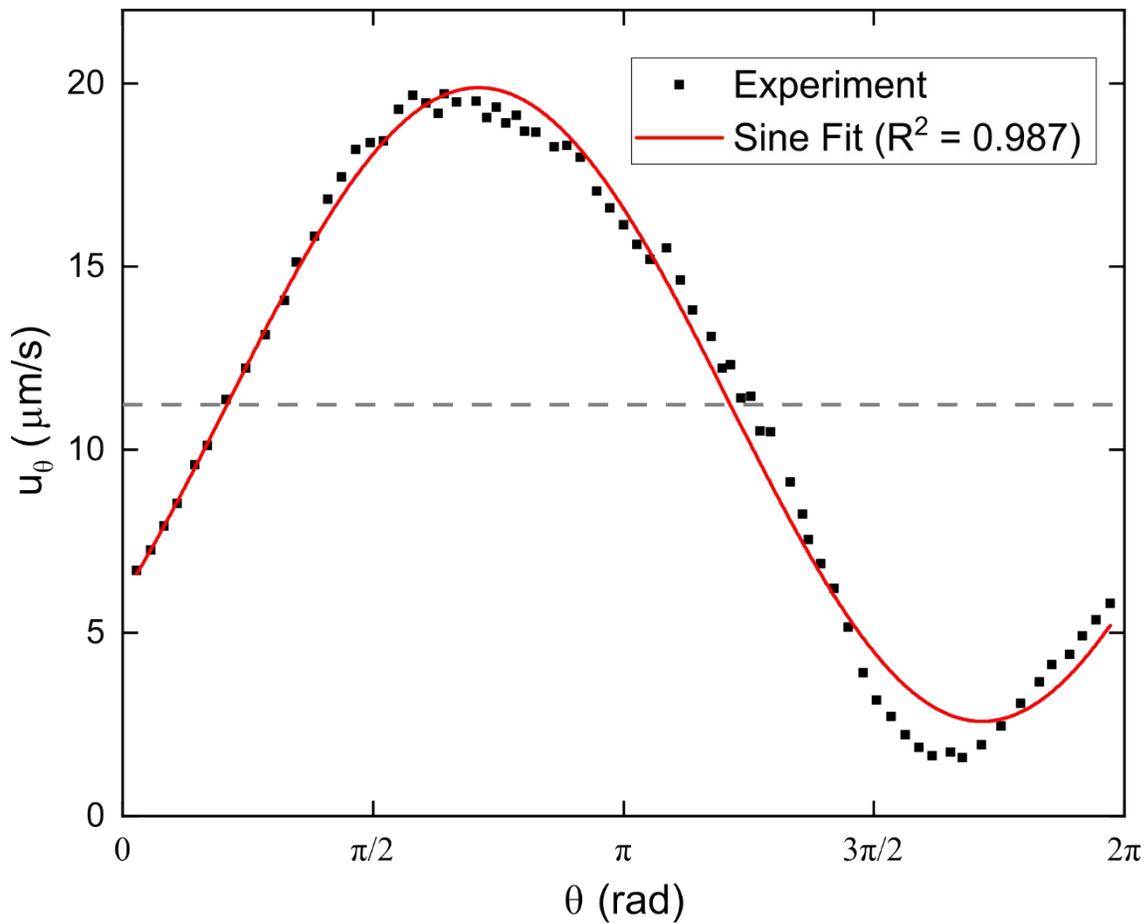

**Supplementary Fig. 11. The plot of tangential velocity ($u_\theta$) vs the angle from x-axis ($\theta$) around a droplet in translating droplet frame during the steering.** The sinusoidal-like profile is due to the Marangoni flows and the positive offset is due to the addition of the short-lived rotational flows caused by the counter-clockwise rotation of the magnetic cluster. The black dots are extracted from experiments and the red curve represents the fitted sinusoidal curve.

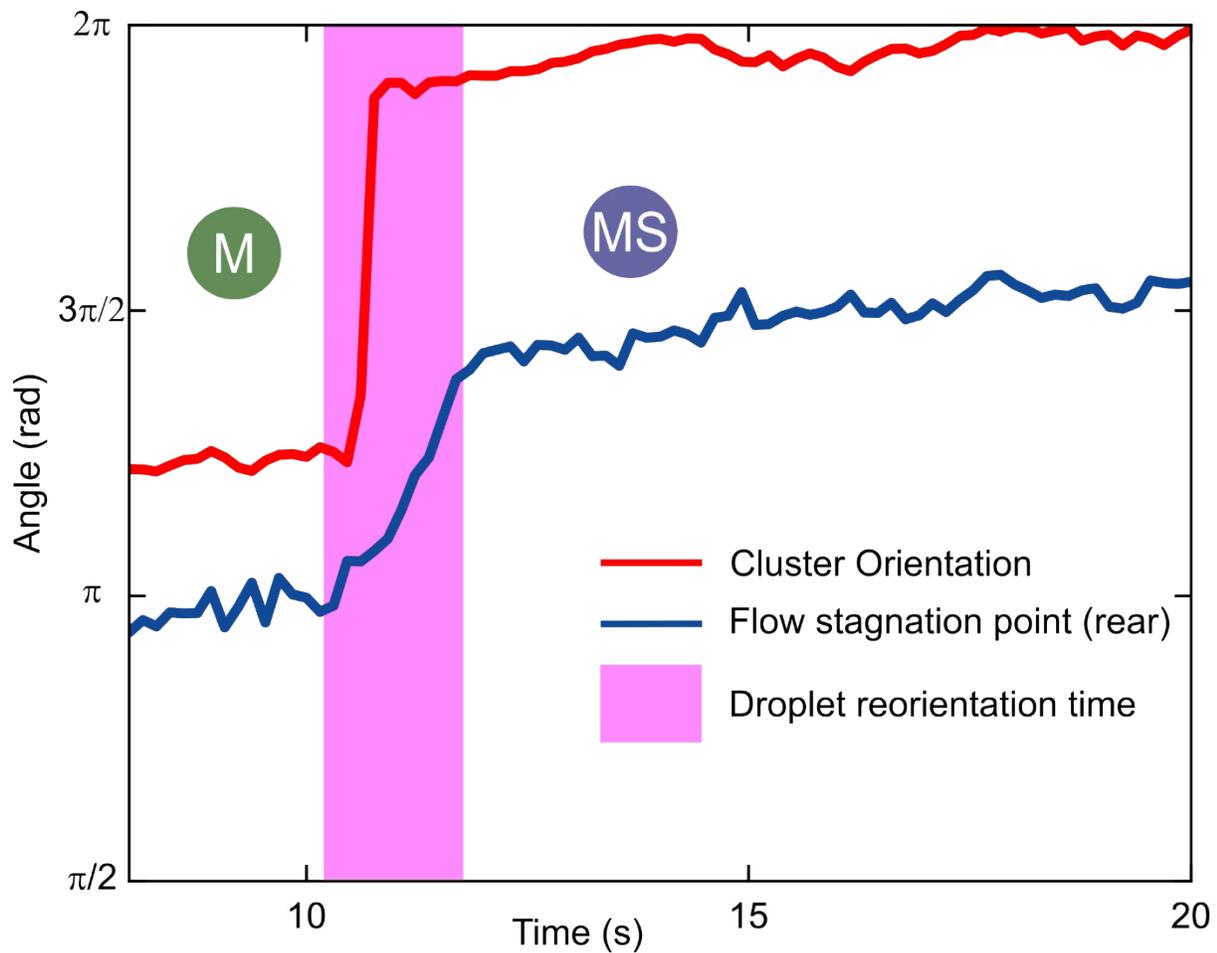

**Supplementary Fig. 12. Comparison of the time scales between magnetic cluster's and droplet's reorientation.** The cluster aligns along the static field in ~0.3 s. The rear stagnation point reorients (corresponding droplet reorientation) in ~1.5 s. The difference in these time scales leads to a finite curvature in the droplet's trajectory.

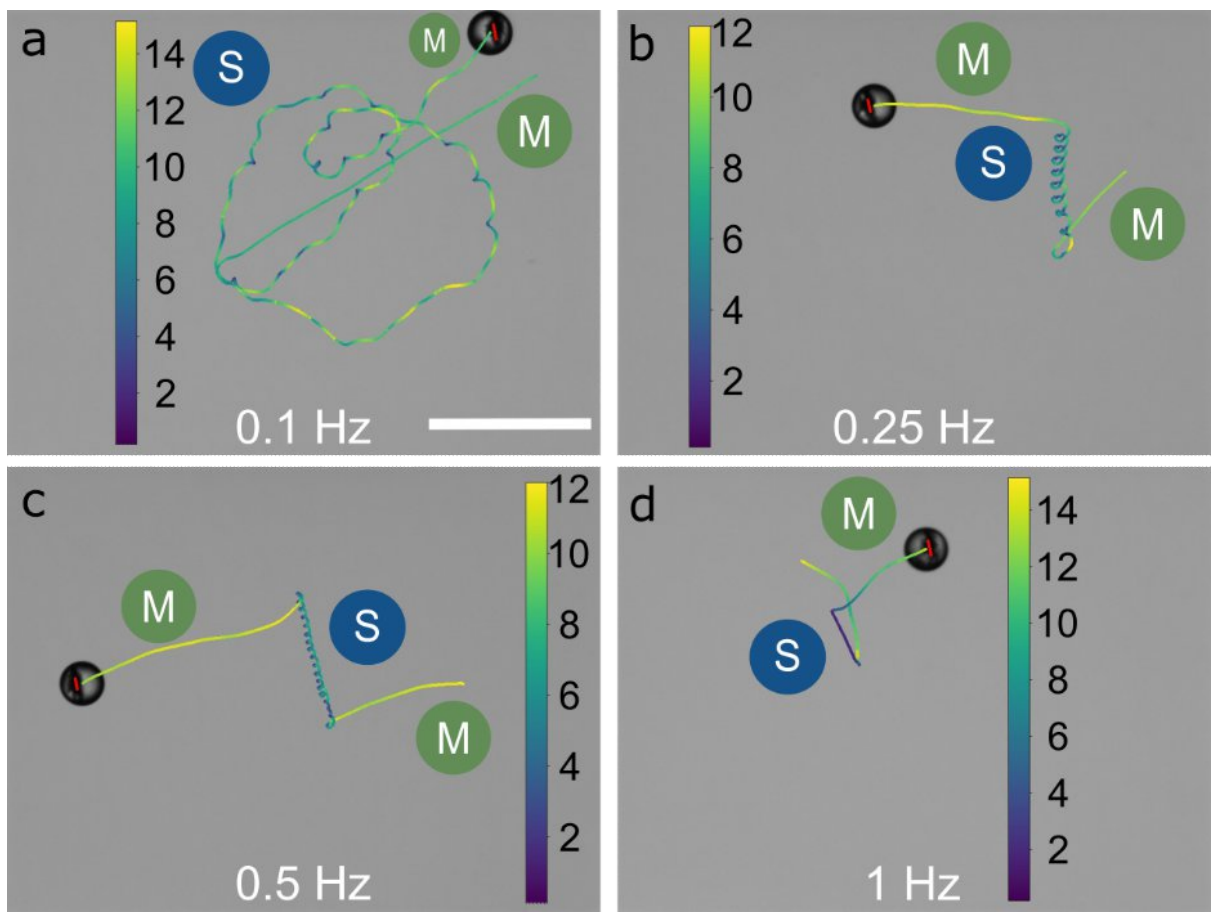

**Supplementary Fig. 13. Switching between curling (S) and straight-line (M) propulsion under in-plane rotating fields at different frequencies.** (**a**) 0.1 Hz. (**b**) 0.25 Hz. (**c**) 0.5 Hz. (**d**) 1 Hz. Trajectories are color coded in instantaneous speed, with color representing speeds in µm s$^{-1}$. Scale bar represents 200 µm.

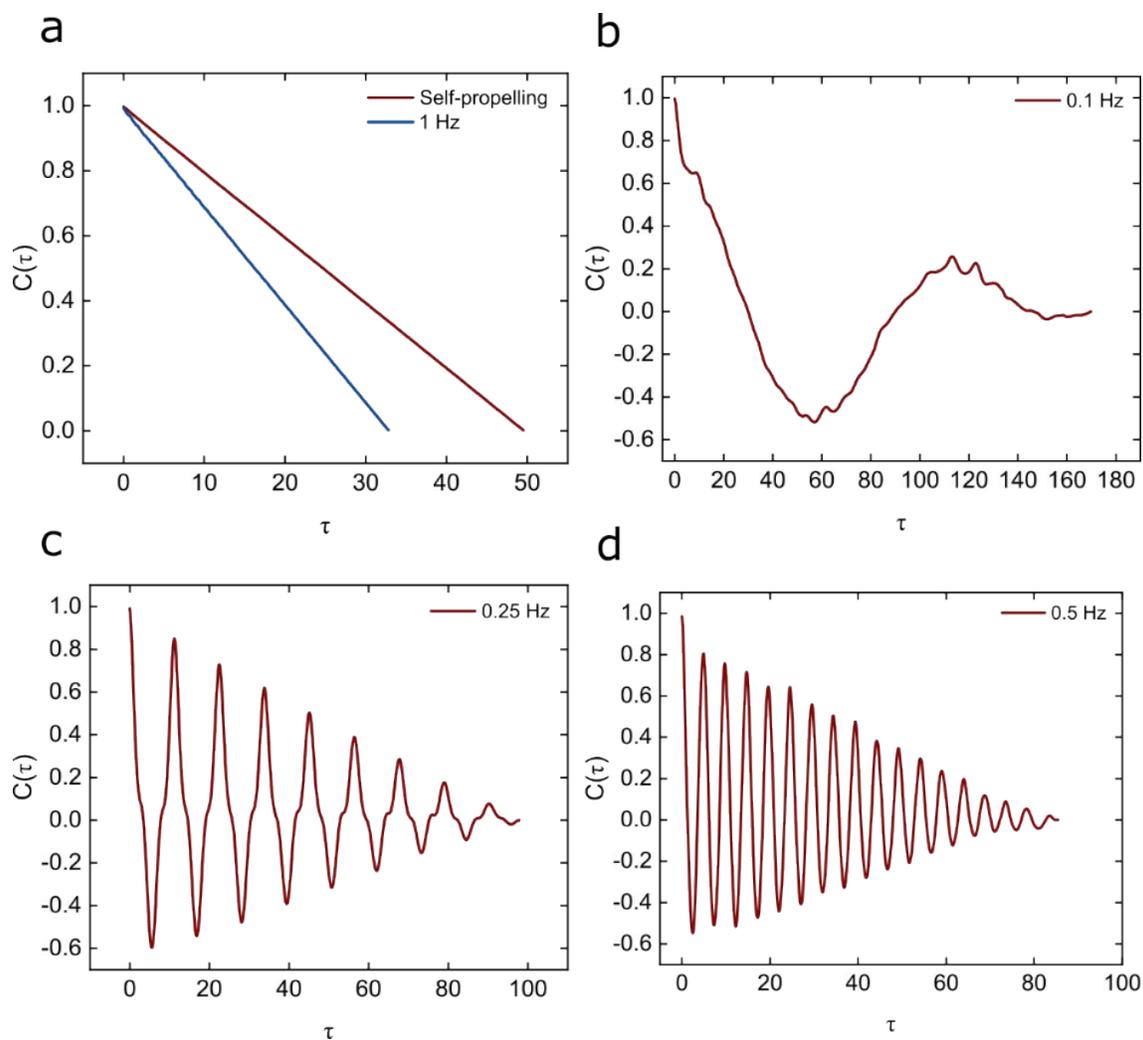

**Supplementary Fig. 14. Angular autocorrelation as a function of time period τ for droplets propelling in straight line (self-propelling) and curling motion at different magnetic field frequencies**. (**a**) Self-propelling (straight line) and 1 Hz. (**b**) 0.1 Hz. (**c**) 0.25 Hz. (**d**) 0.5 Hz. C (τ) decreases over time due to finite length of the velocity signal.

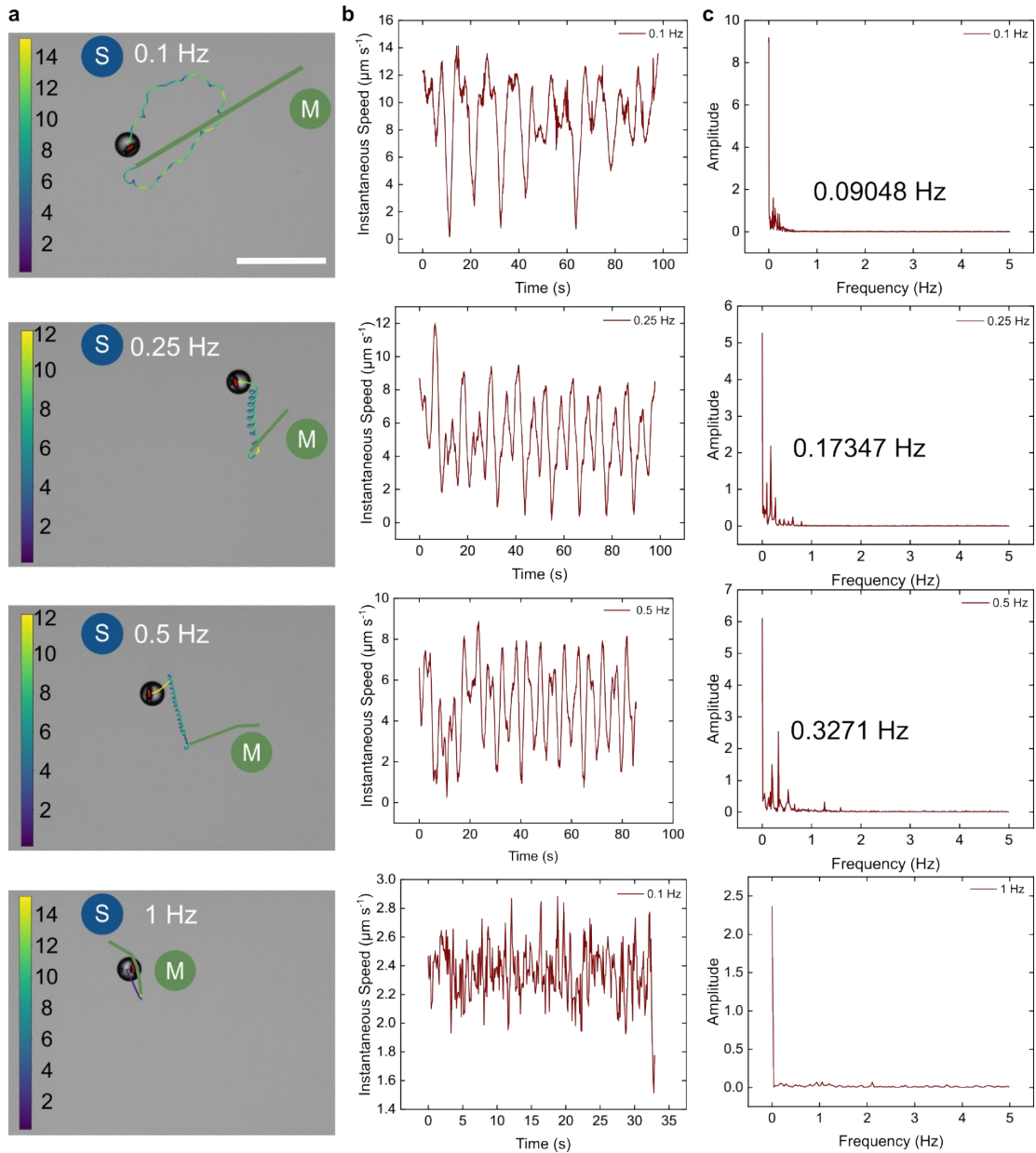

**Supplementary Fig. 15. Speed oscillations during curling motion.** (**a**) Curling trajectories of self-propelling droplets under x-y rotating fields at 0.1 Hz, 0.25 Hz, 0.5 Hz and 1 Hz. Trajectories are color coded in instantaneous speed, with color bar representing speeds in µm s$^{-1}$. Scale bar represents 200 µm. (**b**, **c**) Characterization of instantaneous speed during the curling motion of droplets. Plot showing oscillating instantaneous speed versus time for different frequencies (b) and resulting FFT spectrum with frequency signals (c).

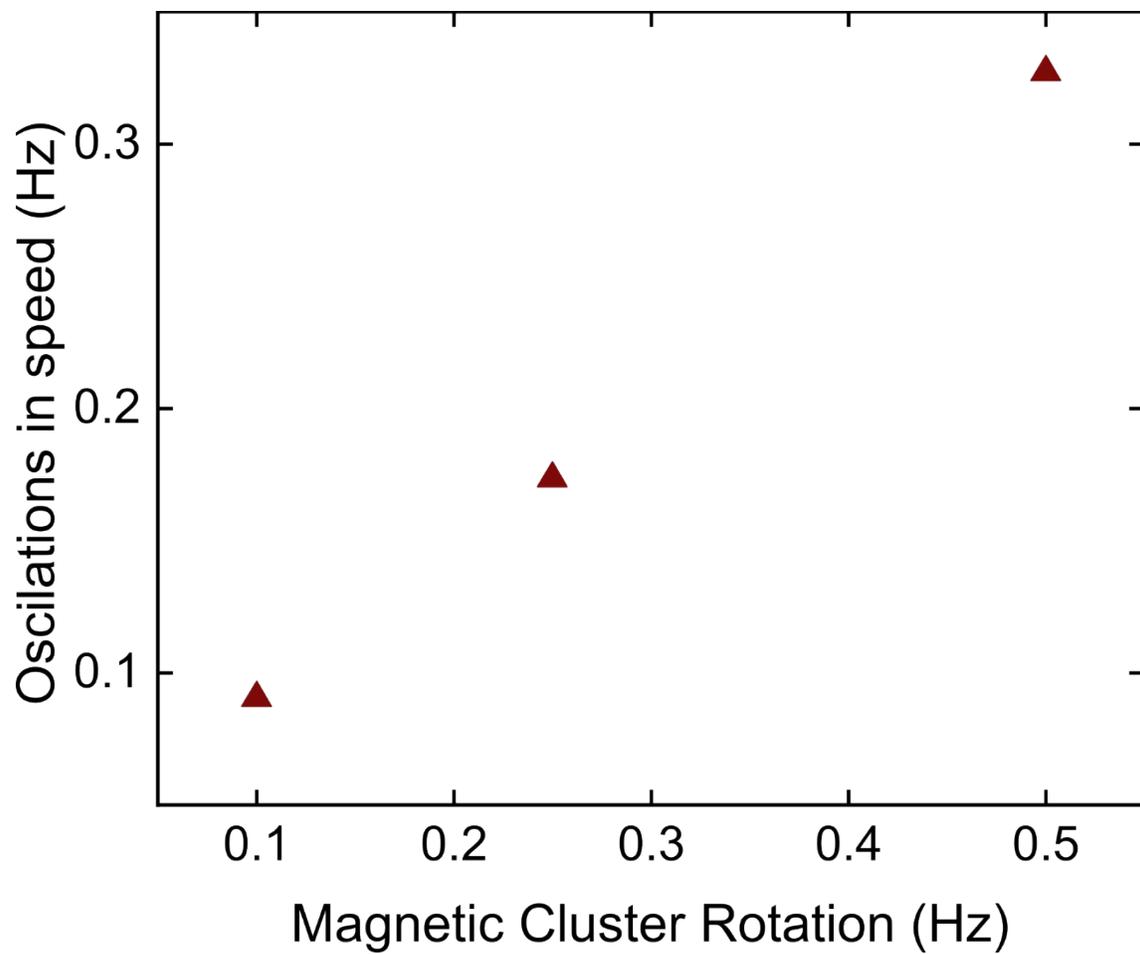

**Supplementary Fig. 16. Frequency of oscillations in instantaneous speeds of droplets undergoing curling motion vary linearly with frequency of magnetic cluster rotation.**

**Supplementary Video captions 1 to 7**

**Supplementary Video 1: Magnetically steering self-propelled droplets using uniform static fields (B = 1 mT).**
First, we observe the behavior of magnetic cluster within a droplet when the droplet starts propelling. An 8CB droplet is heated from 30 °C (smectic state) to 37 °C (nematic state) and the initially stationary droplet starts to self-propel (color bar correspond to instantaneous speed) as the Marangoni flows establish. Correspondingly, we observe that the magnetic cluster translates along the propulsion directions and rotates about its longitudinal axis, while the orientation of the longitudinal axis (denoted by red line) remains bound between ~60-90°. Next, we demonstrate that perturbing the magnetic cluster within a self-propelling droplet (green trajectory) can lead to reorientation of the droplet and it adapts to a new trajectory (purple) perpendicular to the static field direction (in-plane, 1mT). Furthermore, droplets of different emulsion systems (8CB, 5CB, 1-Bromohexadecane and HBA) can be controlled to move in a square trajectory (color bars denote instantaneous speed) by using consecutive 90º magnetic cluster rotations. The red arrows represent the direction of the static magnetic field.

**Supplementary Video 2: Flows around a self-propelling droplet before and after steering.**
Experimental video showing the fluid flows around a propelling droplet. Flows around the droplet are quantified using Particle image velocimetry analysis (PIV). The arrows represent the local fluid velocity around the droplet in the laboratory frame. When the magnetic field is switched on, the magnetic cluster reorients to align along the field and generates short-lived rotational flows. The flow field around the droplet reorients to adapt to the new orientation of the cluster and the droplet propels perpendicular to the applied field.

**Supplementary Video 3: Curling motion under in-plane rotating fields.**
Droplets inherently propel in a straight-line and can be made to propel in curling trajectory by continuously rotating the magnetic cluster using in-plane rotating fields. Resulting helical motion depends on the frequency of rotating magnetic fields.

**Supplementary Video 4: Flows around curling droplets.**
The fluid flows corresponding to the curling motion of three droplets at different rotation frequencies (0.1 Hz, 0.2 Hz and 0.5 Hz) are measured using PIV. The arrows represent local fluid velocity around the droplets in the laboratory frame (top). The flows become more uniform as the rotation frequency increases due to the increasing strength of the rotational flows, as can be seen from the visualization of tangential flow velocities in the droplet frame (bottom). The color bars denote the tangential velocities.

**Supplementary Video 5: Formation and steering of pairs and lines of 8CB droplets.**
Experimental video showing two 8CB droplets forming a hydrodynamic pair due to their inter-droplet interactions. The pair is steered using external magnetic field (color bar denotes time). The strategy is also extended to steer a collective of 8CB droplets that form lines along the magnetic field direction.

**Supplementary Video 6: Controlling collectives of 5CB droplets.**

5CB droplets interact to form aggregates of many droplets. We demonstrate reorientation of the collective using static magnetic field and rotation of the collective under slow rotating magnetic field (0.2 Hz). The trajectory of outermost droplet is color coded with its instantaneous speed and the trajectory of the biggest droplet is color coded with time. The droplet collective splits at higher rotation frequencies (0.5 Hz) and the droplets start to reaggregate when magnetic field is switched off (trajectories color coded with time). Further, higher rotation frequency (5 Hz) is used to split and disperse many aggregates and the droplets slowly reaggregate when the magnetic field is switched off.

**Supplementary Video 7: Controlling predator-prey pairs of magnetic particles-doped BrOCT (predator) and MFB (prey) droplets.**
BrOCT and MFB droplets form a predator-prey pair due to hydrodynamic and oil-exchange interactions. Similarly, a magnetic particle-doped BrOCT also forms a predator-prey pair with a bare MFB droplet. We demonstrate steering, curling motion and splitting of the predator-prey pairs. The trajectories are color coded to distinguish between predator and prey droplets.